\documentclass[conference]{IEEEtran}
\IEEEoverridecommandlockouts
\usepackage{cite}
\usepackage{amsmath,amssymb,amsfonts,amsthm}
\usepackage{enumerate}
\usepackage{algorithmic}
\usepackage{bbm}
\usepackage{graphicx}
\usepackage{textcomp}
\usepackage{xcolor}
\usepackage{subfig}

\theoremstyle{definition}

\theoremstyle{plain}

\newtheorem{thm}{Theorem}

\newtheorem{cor}{Corollary}

\theoremstyle{definition}

\newtheorem{rem}{Remark}

\newcommand{\IEEE}{0}
\newcommand{\TechRep}{1}

\newcommand{\False}{0}

\newcommand{\pubMode}{\TechRep}
\newcommand{\Arxiv}{\False}

\newcommand{\thmSlotCorrelationsFD}{3}
\newcommand{\corNakagamiFD}{2}

\newcommand{\myxmark}{\text{\sffamily \textit{x}}}

\newcommand{\field}[1]{\mathbb{#1}} 
\newcommand{\E}{\field{E}}          
\newcommand{\N}{\field{N}}          
\renewcommand{\Pr}{\field{P}}          
\newcommand{\R}{\field{R}}          

\newcommand {\ones}{\mathbf{1}}

\newcommand{\sra}{\rightarrow}
\newcommand {\abs}[1]{
	\left\lvert #1 \right\rvert 
}
\newcommand {\norm}[1]{
	\left\lVert #1 \right\rVert 
}

\newcommand{\n}[1]{
	\lVert #1 \rVert 
}

\newcommand{\one}{ \mathbf{1}}
\newcommand{\ind}{ \mathbbm{1}} 
\newcommand{\cov}{ \mbox{cov}}

\newcommand{\df}[1]{\textit{#1}}
\newcommand{\ds}{\;\,\;}

\if\pubMode\IEEE
    \newcommand{\comment}[1]{}
\else
    \newcommand{\comment}[1]{\color{red} #1 \color{black} }
\fi

\newcommand{\numberOfSteps}{250}

\newenvironment {rom}{
	\begin{enumerate}[\normalfont(i)]  
	}{
	\end{enumerate}
}
\newenvironment {abc}{
	\begin{enumerate}[\normalfont(a)]  
	}{
	\end{enumerate}
}
\def\BibTeX{{\rm B\kern-.05em{\sc i\kern-.025em b}\kern-.08em
    T\kern-.1667em\lower.7ex\hbox{E}\kern-.125emX}}

\if\pubMode\IEEE
\if\Arxiv\False
\IEEEoverridecommandlockouts
\IEEEpubid{\makebox[\columnwidth]{978-1-7281-4490-0/20/\$31.00 \copyright~2020~IEEE }
\hspace{\columnsep}\makebox[\columnwidth]{ }}
\fi
\fi

\begin{document}
\title{Analytical Derivation of Outage Correlation in Random Media Access with Application to Average Consensus in Wireless Networks
\thanks{This work was funded by the German Research Foundation (DFG) within their priority program SPP 1914 ``Cyber-Physical Networking''.}
}

%
\author{\IEEEauthorblockN{Daniel Schneider}
\IEEEauthorblockA{\textit{Faculty of Computer-Science} \\
\textit{University of Koblenz-Landau}, Germany \\
schneiderd@uni-koblenz.de}
\and
\IEEEauthorblockN{Hannes Frey}
\IEEEauthorblockA{\textit{Faculty of Computer-Science} \\
\textit{University of Koblenz-Landau}, Germany \\
frey@uni-koblenz.de}
}

\if\pubMode\TechRep
 \twocolumn[
  \begin{@twocolumnfalse}

 \large
This  paper  is  a  preprint  (IEEE  “accepted”  status). IEEE copyright notice. \copyright 2015
IEEE. Personal use of this material is permitted. Permission from IEEE must be
obtained for all other uses, in any current or future media, including
reprinting/republishing this material for advertising or promotional purposes,
creating new collective works, for resale or redistribution to servers or lists, or
reuse of any copyrighted.
 \end{@twocolumnfalse}
]
\newpage
\fi

\maketitle

\begin{abstract}
We study a finite and fixed relative formation of possibly mobile wireless networked nodes. The nodes apply average consensus to agree on a common value like the formation's center. 
We assume framed slotted ALOHA based broadcast communication. Our work has two contributions. First, we analyze outage correlation of random media access in wireless networks under Nakagami fading. Second, the correlation terms are applied to the so called L2-joint spectral and numerical radii to analyze convergence speed of average consensus under wireless broadcast communication. This yields a unified framework for studying joint optimization of control and network parameters for consensus subject to message losses in wireless communications. Exemplary we show in this work how far outage correlation in wireless broadcast communication positively affects convergence speed of average consensus compared to consensus in the uncorrelated case.

\end{abstract}
\section{Introduction}
During the past decades wireless sensor networks (WSN) and distributed multiagent systems have gained a lot of attention among both communication and control community.
While there is much work on the specifics of either side, only few analytical papers consider a unifying approach where the impact of real wireless network phenomenona is measured directly in terms of distributed control performance. 

In this work, we follow such unifying approach. At first, we quantify outage correlations due to random media access control for finite WSN's and for snapshots of multi agent systems. Thereafter, we consider distributed average consensus, a basic building block of distributed control,
and lay open that a thorough analysis of control performance requires knowledge of how packet losses are correlated.

When speaking of correlation, we mean outage correlation \cite{Talarico2018} in decorrelated fading channels (communication channels are typically located such that they are neither spatially nor temporally correlated).
The correlation we study here comes from random media access.


We consider snapshots of node deployments 
and identify two components of correlation. First, correlation between all recipients of the same transmitter, and second, correlation between all other links. Typically, the former is positive, the latter is mostly negative.

The applications we have in mind all require broadcast transmission such that Request To Send/Clear To Send mechanisms cannot be applied. Examples consist of formation control \cite{Dutta2016}, flocking (cf. the references in \cite{Olfati-Saber2007}) 
or purely computational tasks like determining the network size \cite{Sluciak2013}. Literature in the former field
have been considering 
oversimplified wireless network assumptions so far \cite{Dutta2016}, \cite{Sardellitti2012}.

We build our analysis on a framed slotted ALOHA media access protocol and measure the effects 
on the performance of an average consensus protocol. 
The simplicity of slotted ALOHA allows for exact theoretical analysis of the control performance for the given deployment. 
Moreover, even if carrier sensing was applied, it would still become less and less beneficial with increasing node density.
For example, it was theoretically studied in \cite{Baccelli13} that the IEEE 802.11p protocol behaves like slotted ALOHA in the dense regime.


The remainder of this work is structured as follows. In the next section we relate our study to existing work on beneficial and disadvantageous effects of correlation on communication and control performance. In section~\ref{sec:definitions} we introduce mathematical notation and the physical layer model under consideration. This is followed by section~\ref{sec:corr} where we derive an expression to compute outage correlation of random media access given an underlying fading model. This result is then used in section~\ref{sec:application} where we describe the effects correlation implies for consensus. We provide an exact and narrow performance region utilizing the so called $L^2$-joint spectral radius ($r_2$), e.g. \cite{Ogura2013}, and $L^2$-joint numerical radius ($w_2$) \cite{ICCCN19} for mean square analysis on the convergence speed of average consensus. Both measures together quite precisely render the outage correlation in terms of (discrete) control gain, fading parameter and frame length.\ 
Compared to an uncorrelated model with the same packet loss rates, we essentially observe a beneficial impact of correlation 
on average consensus performance. We conclude our findings in section~\ref{sec:conclusion}. 

\section{Related work}
The performance impact of correlated communication channels is studied in an extensive field of applications. It is known that correlation in general can have positive effects, e.g. on multiple-input and multiple-output capacity \cite{Benef04} or for acknowledgements \cite{Alam2012}, or drawbacks, for instance for retransmissions \cite{Ganti2009a}.
More specifically, other correlation types can be found which we do not consider or are ruled out in this work, e.g.
correlation between different antennas or temporal correlation \cite{Haenggi14}.

In this work, our focus is on outage correlation, that is the correlation between packet losses along different links. In the chosen slotted ALOHA media access scheme, also other types of correlation arise which implicitly affect the outage correlation and which may be subsumed under the term interference correlation.
This includes slot correlation (i.e. correlation between different slots within the same frame), spatial correlation and correlation between different receivers, cf.\cite{Haenggi14}.


Interference correlation can be studied for fixed node configurations or for networks 
modeled as point processes as studied in \cite{Ganti2009a}, \cite{Haenggi14}, \cite{Krishnan2017}. 
Recently, outage correlation has been investigated in a finite point process setting (Binomial, Poisson and Thomas processes) under Rayleigh Fading \cite{Talarico2018}. 
Although the authors allow for slotted ALOHA, they are mainly concerned with the effects of the spatial differences generated by the three point processes; concrete node deployments are not considered. 
The work \cite{Kim2019} is set also in a point process setting where spatial interference correlation is investigated. Instead of computing the correlations exactly, the authors elaborate approximations. 

Our analytical derivations significantly differ from the aforementioned works. Here we are specifically interested in the impact of outage correlation for a finite, fixed node deployment with respect to convergence speed of average consensus using broadcast communication.

For convergence speed of average consensus under channel correlation and wireless network constraints other work exist. In\cite{Chan10}, consensus has been considered under fast fading where correlated outages (mainly due to fading) are mentioned without being quantified. In \cite{Kar2009}, \cite{Silva11} abstract correlations can be handled in the consensus framework, however, neither a physical model nor a media access model is provided. Also the mean square disagreement measures used there are upper bounds only. In our work we use a different upper bound, $w_2$, which very well matches the shape of the exact measure $r_2$ characterizing (exponential) mean square stability, cf. \cite{ICCCN19}. 

Finally, the outage probabilities for Nakagami-m fading we plug into our model (Corollary~\ref{cor:nakagamiHD})
stem from \cite{Nakagami12}, one of the few papers analyzing fixed node deployments. However, as opposed to our work, correlation is not considered in that reference.

\section{Definitions and model assumptions}\label{sec:definitions}
\subsection{Matrix, vector and other notation}
\label{notation_section}
For vectors $x\in \R^n$ we adopt the Euclicean norm $\norm{x} = \sqrt{\sum_{i=1}^{n} x_i^2}.$ By $\one \in \R^n$ we denote the vector of all ones. The identity matrix on $\R^n$ is denoted by $I_n,$ whereas by $\Pi$ we denote the projection matrix $I_n - \frac{\one \one^T}{n}.$
%

The Kronecker delta is denoted by $\delta_{ij}$, i.e. $\delta_{ij} = 1$ if $i = j$ and $\delta_{ij} = 0$ otherwise; $\delta_{i\ne j}$ is short for $1-\delta_{ij}.$  For two matrices $A,B \in \R^{n\times n}$ we denote by $A\otimes B$ the Kronecker product of $A$ and $B,$ i.e. for indices of the form $I = n\cdot (i-1)+k,\; J=n\cdot (j-1) +l,\; 1 \le i,j,k,l \le n,$ we have $(A\otimes B)_{IJ} = a_{ij}b_{kl}.$ 

By $\ind(x\ge0)$ we denote the unit step function on $\R$, i.e. $\ind(x\ge0)$ is equal to one if $x\ge0,$ 
and vanishes otherwise.

Let $X$ be an arbitrary set and $n\in \N.$ By $X^n$ we denote the $n$-fold cartesian product $X \times \dots \times X.$ 
We will use ordered tuples over sets $X$ 
as follows: For a sequence $(a_\nu)_{\nu \in \N} \subset X$ and a finite index set $J \subset \N$ we denote by $[a_\nu]_{\nu \in J}$ the tuple $(a_{\nu_{1}}, \dots, a_{\nu_{\abs{J}}}) \in X^{\abs{J}}$ where 
$\nu_1 = \min J, \mbox{ and }  \nu_{k+1} = \min J\setminus\{\nu_1, \dots, \nu_k\}, \ds k = 1, \dots \abs{J} -1.$

\subsection{Physical layer model}
We assume $n \ge 3$ nodes $u_1, \dots, u_n$ located at fixed positions $(x_1, \dots, x_n) \in \R^{3n}$ relative to a possibly moving frame of reference. A transmission between two nodes $u_i$ and $u_j$ is subject to path loss with a path loss coefficient of $\alpha_{\text{PL}} \ge 2$. For a transmission distance $d$ the average received power is thus proportional to $1/d^{\alpha_{\text{PL}}}$
in the far field. 

Let $\mu_{ij}^{-1}, \, 1 \le i, j \le n,\; i \ne j,$ be the average power of a transmission from $u_i$ received at $u_j$.
If $u_i$ transmits with power $q_i,$ we have 
$\mu_{ij} = \frac{1}{q_{i}}(\norm{x_i - x_j}/r_0)^{\alpha_{\text{PL}}}$, where $r_0$ is a reference distance. Denoting the wavelength of the carrier wave by $\lambda$ we choose $r_0 = \lambda/4\pi$ such that we arrive at Friis free space equation in case of a path loss coefficient of $2.$

Here we study communication under the effect of narrow band fast fading. This means that the instantaneous received power $P_{ij}$ of a transmission between $u_i$ and $u_j$ is a non-negative random variable with expectation $\mu_{ij}^{-1}$. 
Further, we assume independent block fading (IBF), i.e. $P_{ij}$ is constant during a packet transmission and power variables belonging to different channels are mutually independent.
Later we employ the Nakagami fading model, i.e., the random variable $P_{ij}$ is Gamma distributed with probability densitiy function (pdf) 
\begin{equation}
	\label{eqn:naka_pdf}
	f_{ij}(x) = \ind(x\ge0) \frac {m_{i}^{m_{i}}\mu_{ij}^{m_{i}}}{\Gamma (m_i)} x^{m_{i}-1}\exp \left(-m_{i}x\mu_{ij} \right)
\end{equation}
where $m_i  \in \N$ are integer\footnote{In general, Nakagami fading only requires $m_i \ge 1/2,$ however since we build on \cite{Nakagami12}, we can only handle integer values.} shape parameters. 
However, our calculations of the structure of correlation due to random media access are valid in a very general setting regardless of path loss and fading (cf. state coefficients of Theorem~\ref{prp:slotCorrelationsHD} and 
\if\pubMode\IEEE
     \cite[Theorem~\thmSlotCorrelationsFD{}]{TechRep}
\else
    \ref{prp:slotCorrelationsFD}
\fi).
We incorporate Nakagami fading as special case in Corollary~\ref{cor:nakagamiHD} and 
\if\pubMode\IEEE
     \cite[Corollary~\corNakagamiFD]{TechRep}
\else
    \ref{cor:nakagamiFD}.
\fi.

The success rate of a message transmission is modeled with outage probability under the signal to interference plus noise ratio (SINR) model. For a transmission from $u_i$ to $u_j$ all other currently ongoing transmissions will be considered as noise. Thus, we relate the received power over noise plus all other interfering transmissions as
\begin{equation}
	\label{eqn:SINR}
	\mathrm{SINR}_{ij} = \frac{P_{ij}}{N + \sum_{\nu \in V \setminus \{ i,j \}} P_{\nu j}}
\end{equation}
where $N$ is a constant thermal noise term and $V$ the set of node indices of all currently transmitting nodes. Let the receiving node $u_j$ be silent. Then we say the transmission has an outage if $\mathrm{SINR}_{ij}$ falls below a system specific \df{threshold} value $\theta > 0$ and the probability of a successful packet transmission from $u_i$ to $u_j$ is $\Pr[\mathrm{SINR_{ij}} \geq \theta].$
The case when $u_j$ is also transmitting will be discussed in the context of the slot model in the next section.

\color{black}

\section{Outage correlation due to media access}
\label{sec:corr}
To determine the coefficient of correlation of transmission success between different node pairs, it suffices to specify expectation and covariance, which is the concern of this section.


 
In addition to the physical layer model, we now assume that the nodes perform the following variant of slotted ALOHA for broadcast transmission which guarantees that each node transmits exactly once during a beacon period. Let $\tau$ be the duration of the beacon time interval in milliseconds (ms) which is called a \textit{frame.} 
Each frame is divided into a number $m$ of time intervals called \textit{slots}, where $m \in \N$ is chosen such that each packet can be transmitted in $\tau/m$ ms. At the beginning of each frame, each node randomly chooses exactly one slot with equal probability. Transmission from node $u_i$ to node $u_j$ is successful if it is successful in the chosen slot. Otherwise the packet from $u_i$ to $u_j$ is considered lost in this beacon interval.


Our derivation requires\footnote{Cf. Remark \ref{rem:thm} and Section \ref{sec:parameters}. The assumption $\theta \ge 1$ is commonly used in singular antenna systems, \cite{Clazzer2017}. } $\theta \ge 1$ and $N>0$ Watt. Let us further introduce $n$ independent random variables $S_1,\dots, S_n$ representing each node's choice of slot - all equally likely - i.e. $S_i \sim unif(\{1, \dots, m\}), \; i = 1 ,\dots n,$ (uniform distribution).

Moreover we assume the power variables $P_{ij}$ - e.g. $P_{ij} \sim f_{ij},$ cf. (\ref{eqn:naka_pdf}) - to be independent (IBF assumption) and also to be independent from the slot variables. We can now express power and interference in slot $k, \; 1 \le k \le m:$
	$$P_{ij}^{k} := \delta_{S_i k} P_{ij} \ds \mbox{ and }\ds I_{ij}^k := \sum_{\substack{l=1 \\ l\ne i,j}}^{n}P_{lj}^{k}.$$
Let us introduce random variables for excess of the SINR threshold in slot $k$ and in any slot respectively:
	$$X_{ij}^k := \ind\left(\frac{P_{ij}^k}{N + I_{ij}^k}\ge \theta  \right) \ds \mbox{ and } \ds X_{ij} := \sum_{k=1}^{m} X_{ij}^k.$$
	Clearly, from the definition of $P_{ij}^k$ and the fact that $\theta, (N + I_{ij}^k) >0$ we have $X_{ij}^k = \delta_{S_i k} \cdot X_{ij}^k,$ thus $X_{ij} = X_{ij}^{S_i}.$

	There are two further practial aspects we have to consider: Sending nodes cannot receive (singular antenna case) and multiple receptions at some node at the same time (i.e. in the same slot) are impossible in our model since $\theta \ge 1.$  
	We call the former aspect the 
	\df{half-duplex} case. 
%
We define successful \df{full-duplex} packet transmission from node $u_i$ to $u_j$ to be the event $X_{ij} = 1.$ In words: We ignore whether the receiver is sending or not when checking for excess of the SINR ratio.
	
	In this sense, we model successful half-duplex packet reception $Y_{ij}, \, 1 \le i, j \le n,\; i \ne j,$ as Bernoulli variable having the property to vanish if $u_i$ and $u_j$ chose the same slot and to coincide with $X_{ij}$ otherwise:
\begin{equation} \label{eqn:X_Y}Y_{ij} := (1-\delta_{S_i S_j}) \cdot X_{ij} \ds\ds 1 \le i,j \le n, i\ne j.\end{equation}
Our goal is to determine how successful transmissions are correlated. Therefore, we calculate the covariances $\cov(Y_{ij}, Y_{kl})$ and begin with the mixed moments of the $X_{ij}$'s (full-duplex case) which are of special importance.
By the law of total probability we can decompose the mixed moments $\E X_{ij}X_{kl},\, i \ne k,$ into a weighted sum of the conditioned mixed moments
	\begin{equation}\label{eqn:decomposition1_mixed_moments}\frac{1}{m} \cdot \E[X_{ij}X_{kl}\vert S_{i} = S_{k}] + (1-\frac{1}{m})\E[X_{ij}X_{kl}\vert S_{i} \ne S_{k}]\end{equation}
motivating the form of the main result of this section which needs a little technical preparation. The conditioned mixed moments of the $X_{ij}$'s, as well as those of the $Y_{ij}$'s, lead - up to normalization - to expressions of the form (\ref{eqn:formulaSameSlot}) and (\ref{eqn:formulaDiffSlot}) which we will now describe:
Denote by $b_\nu \in \left\{ 1,\dots, m \right\}$ realizations of the slot random variables $S_\nu$ which we call \df{micro states}. Consider for instance a transmission from node $u_i$ to $u_j.$ 
The pattern of the micro states contributing to the interference for this transmission can be summerized by
simpler \df{macro state} variables $c_\nu \in \{0,1,2\}, \nu \ne i,j.$ 
Consider, e.g., a further node $u_k$ under the condition $S_i \ne S_k,$ i.e. $u_i$ and $u_k$ have chosen different slots. In one possible interpretation $c_\nu = 0$ may stand for all realizations of the event $S_\nu \ne S_i, S_k,$ the case $ c_\nu = 1$ might stand for all realizations of $S_\nu  = S_i$ and $c_\nu = 2$ for all realizations of $S_\nu = S_k.$
In general, for a subset $J$ of $\mathcal N:= \left\{ 1, \dots, m \right\}$ we denote by $c$ the tuple $[c_\nu]_{\nu \in J},$ the \df{total (macro) state}. We express the number of micro states represented by this total state by functions $\phi(c)$ and $\psi(c)$ which we call \df{state weights}.

Now let us take a look at some potential receiver $u_l$ of $u_k$'s transmission. The idea in the calculation of ((un-)conditioned) mixed moments of $X_{ij}$ and $X_{kl}$ is to condition on the slot variables until all $S_{\nu}$ have been replaced by micro states $b_\nu.$ The mixed moments then expand into a sum of weighted probabilities. When exploiting common interference terms, the probability summands split up into a product of two separate probabilities in virtue of the  independence
(IBF assumption). For either of the two factors we use the terminology \df{state coefficient} - symbolized by either $\alpha_{ijkl}(c)$ or $\beta_{ijkl}^{(d)}(c).$ These probabilities depend, of course, on the fading model (cf. Corollary~\ref{cor:nakagamiHD}). State weight and state coefficients are also called \df{state functions}. 

For $i \ne j, k \ne l, (i,j) \ne (k,l),$ and for a finite index set $J \subset \mathcal N$ 
we now introduce expressions allowing for a quite general treatment of outage correlation expressions\footnote{The summation appearing in (\ref{eqn:formulaSameSlot}) and (\ref{eqn:formulaDiffSlot}) is meant to expand into a $\abs{J}$-fold sum over $c_\nu,\, \nu \in J,$ e.g. $\sum_{\substack{c_\nu=0 \\\nu \in \mathcal N}}^{1}(\dots) = \sum_{\nu_1, \dots, \nu_n = 0}^1(\dots)$.}
\begin{equation}\begin{aligned}\label{eqn:formulaSameSlot}
		\Phi_{ijkl}^{J}(\phi, \alpha):=  \sum_{\substack{c_\nu=0\\ \nu\in J}}^{1}\phi(c)\cdot \alpha_{ijkl}(c)\cdot \alpha_{klij}(c)
 \end{aligned}\end{equation}
and
\begin{equation}
	\label{eqn:formulaDiffSlot}
	\Psi_{ijkl}^J(\psi, \beta) :=\sum_{\substack{c_\nu=0\\ \nu \in J}}^{2}  \psi(c)\cdot \beta_{ijkl}^{(1)}(c) \cdot \beta_{klij}^{(2)}(c),
\end{equation}
where $\phi(c), \psi(c), \alpha_{xyzw}(c)$ and $\beta_{xyzw}^{(d)}(c)$ are functions of 
$c = [c_\nu]_{\nu \in J} \in   \left\{ 0,1,2 \right\}^{\abs J}$ (see section \ref{notation_section}).
E.g. for $J = \mathcal N \setminus \left\{ i,k \right\}$ we have $[c_\nu]_{\nu \in J} = (c_1, \dots, c_{\mu-1},c_{\mu+1},\dots, c_{\hat \mu-1}, c_{\hat \mu+1},\dots, c_n),$ where $\mu = \mbox{min}(i,k)$ and $\hat \mu = \mbox{max}(i,k).$ 
In case of the calculation of $\Phi_{ijkl}^J$ we always have $c_\nu \ne 2$ such that there $[c_\nu]_{\nu \in J} \in   \left\{ 0,1 \right\}^{\abs J}$ holds true.

\begin{figure}[t!]
	\centerline{\includegraphics[trim=400 80 400 40,clip, width = 0.35 \textwidth]{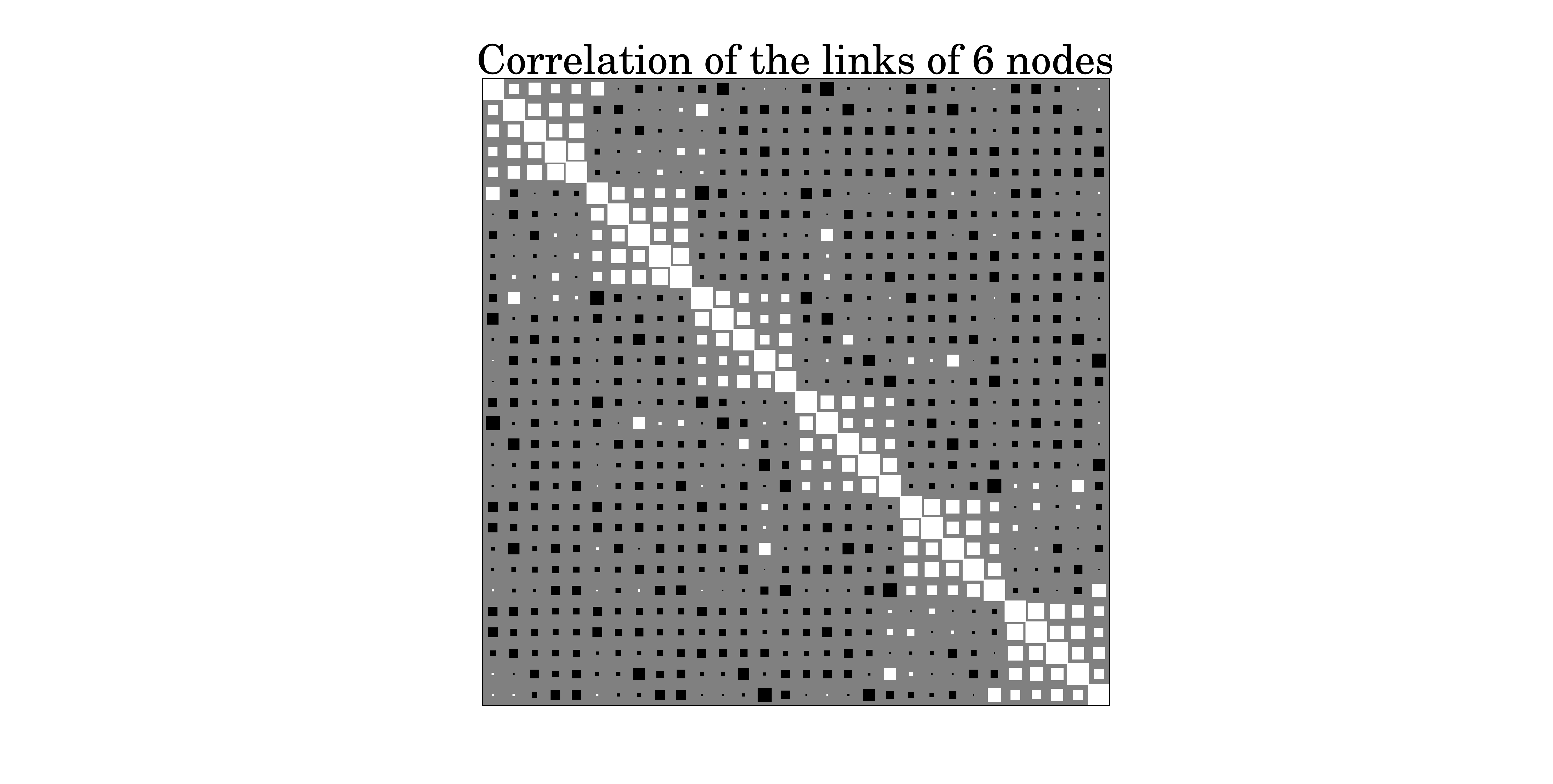}}
	\caption{\scriptsize Distribution of the correlation for a half-duplex example. White and black squares correspond to positive and negative correlation, respectively. A larger square means a larger correlation in modulus. At the largest ones on the diagonal the correlation is $+1.$ 
	The component (a) of Theorem \ref{prp:slotCorrelationsHD} corresponds to the white diagonal square blocks. \label{fig:correlation}}
\end{figure}

We are now ready to formulate our covariance result for the $Y_{ij}$'s. An analogous result for the $X_{ij}$'s can be found in 
\if\pubMode\IEEE
the Appendix of \cite{TechRep}.
\else
Appendix~\ref{sec:full_duplex}.
\fi
From (\ref{eqn:X_Y}) we obtain for $i\ne k$ and $k\ne l$

	$$ \begin{aligned}\E [Y_{ij} Y_{kl}] = \left(1-\frac{1}{m}\right)^{2-\delta_{kj}\delta_{li}} \E [X_{ij} X_{kl} \mid S_i \ne S_j; S_k \ne S_l]. \end{aligned}$$
In the following Theorem we treat the cases $S_i = S_k$ and $S_i \ne S_k$ separately,  cf. (\ref{eqn:decomposition1_mixed_moments}).
\begin{thm}[Half-duplex covariances]
\label{prp:slotCorrelationsHD}
\item Let $n \ge 3, \theta \ge 1,N > 0, i \ne j, k \ne l,$ and let $(i,j) \ne (k,l).$ (a)\; We have for $j\ne k$ and $i\ne l$

$$  \begin{aligned} \E &\left[ X_{ij}X_{kl} \mid S_i = S_k; S_i \ne S_j; S_k \ne S_l \right] \\&= \frac{ (1-\delta_{l j})\cdot \Phi_{ijkl}^{\mathcal N \setminus\left\{ i,j,k,l \right\}}(\phi, \alpha)}{m^{n-4+\delta_{ik}}},  \end{aligned} $$ 

where the state weight of $c = [c_\nu]_{\nu\in \mathcal N\setminus \left\{ i,j,k,l \right\}}$ is given by
	\begin{equation}\label{eqn:state_weight_a_HD}\phi(c) = \prod_{\nu \in \mathcal N \setminus \left\{ i,j,k,l \right\}}(m-1)^{1-c_\nu},\end{equation}
	the state coefficients $\alpha_{xyzw}(c)$ are given by 
 	\begin{equation*}
		\begin{aligned}
		\Pr\left(\frac{P_{xy}}{N+\sum\limits_{\nu\ne x,y,z,w}^{}c_\nu P_{\nu y} + (1-\delta_{xz})P_{zy}} \ge \theta\right).		\end{aligned}
	\end{equation*} 
In particular, $\E[ Y_{ij}Y_{il}]$ can be expressed this way, hence for $j \ne l$
	\begin{equation}\label{eqn:cov_a_HD}cov(Y_{ij},Y_{il}) = \left(1-\frac{1}{m}\right)^2\cdot \frac{\Phi_{ijil}^{\mathcal N \setminus\left\{ i,j,l \right\}}}{m^{n-3}}(\phi, \alpha) - \E [Y_{ij}]\E[ Y_{il}].\end{equation}
(b) Let $J' = \{i,j,k,l\}$, let $J'' = \{i,k\}$ if $\abs{J'} =4$ and $J''=J'$ else. Moreover, let $J'''=\{i,j,k\}$ if $\abs{J'}=4$ and $J''' = J'$ otherwise. Then, for $k\ne i,$ 
we have, 
	$$ \begin{aligned}&\E [X_{ij}X_{kl}\vert S_i \ne S_j, S_k;\, S_k \ne S_l]  = \frac{\Psi_{ijkl}^{\mathcal N \setminus J''}(\psi, \beta)}{M}
			,\end{aligned}$$
		where $M = m^{n-\abs{J'}}$ if $\abs{J'} = 2, 3$  and $M=m^{n-3}\cdot(m-1)$ if $\abs{J'}=4$. Moreover, for $k \ne i,$
        \begin{equation}\label{eqn:cov_b_HD}\begin{aligned} &\cov(Y_{ij}, Y_{kl}) = 
		(1-\frac{1}{m})^{2-\delta_{kj}\delta_{li}}\left[ \frac{\delta_{\abs{J'},4} (m-1)\Phi_{ijkl}^{\mathcal N \setminus J'}(\phi, \alpha)}{m^{n-3}}\right.\\
			&   +\left. 
            \left[\frac{m-1}{m} \delta_{\abs{J'},4} + \frac{m-2}{m-1}\delta_{l,j} +
            \delta_{l\ne j}\delta_{\abs{J'}\ne 4}\right]
			\frac{\Psi_{ijkl}^{\mathcal N \setminus J''}(\psi, \beta)}{M}\right] \\
			& - \E [Y_{ij}]\cdot \E [Y_{kl}], 
			\end{aligned}\end{equation}
	    where $\Phi_{ijkl}(\phi, \alpha)$ depends on the state functions of part (a) whence $\Psi_{ijkl}(\psi, \beta)$ depends on the state weight
	    \begin{equation}\label{eqn:state_weight_b_HD}\psi(c) =  \left[\delta_{c_l\ne 2}\cdot  \delta_{c_j\ne 1} \cdot( m-1)^{\delta_{0,c_j}}\right]^{\delta_{\abs{J'},4}} \prod_{\substack{\nu = 1\\ \nu \not\in J'''}}^n (m-2)^{\delta_{c_\nu, 0}}\end{equation}
	and on the state coefficients
		$$\beta_{xyzw}^{(d)}(c)  = \Pr\left(\frac{P_{xy}}{N+\sum_{\nu\ne x,y,z}^{}\delta_{c_\nu d} P_{\nu y}} \ge \theta  \right),$$
where $c = [c_\nu]_{\nu\in \mathcal N\setminus J''}$,  $d = 1,2.$ 
 
\end{thm} 
\begin{proof}
\if\pubMode\IEEE
    The proof utilizes the conditioning and macro state techniques sketched above. For details see \cite[Appendix]{TechRep}.
\else
    See Appendix \ref{prf:slotCorrelationsHD}.
\fi
\end{proof}
\begin{rem}
	\label{rem:thm}
	\begin{rom}
	\item The requirement $\theta \ge 1$ is used in the case $l = j$ and yields the $(1-\delta_{lj})$ factor.
	\item In case $m=1$ there is no outage correlation in the full-duplex case, i.e. $\cov(X_{ij}, X_{kl}) = 0$ for $(i,j) \ne (k,l).$ 
	\item Typically, the covariance component given by (a) seems to be stronger than the one given by (b). 
	For the correlations this observation does not hold; we may merely assert that the component due to (a) is typically positively correlated while the one due to (b) is correlated mainly negatively, 
	cf. Figure~\ref{fig:correlation}.
	\item Since the summations can get large it is important to efficiently implement the sums. E.g. $\sum_{\substack{c_\nu = 0\\ \nu \ne i,k}}^1 \left( \dots \right)$ can be implemented using a loop from 1 to $2^{n-2}$ over a \textit{single} integer variable $i$ and accessing the ``components'' $c_{\nu}$ of $i$ by bitwise operations (bit-shift).
 
	\end{rom}
\end{rem} 

\begin{rem}
\label{rem:backward_compatibility}
Using Rayleigh fading in the slot model we recover the success probabilities of \cite{Nakagami12} (full-duplex) and \cite{ICCCN19} (half-duplex). The calculation is provided in 
\if\pubMode\IEEE
    \cite{TechRep}.
\else
    Appendix \ref{prf:backward_compatibility}.
\fi
Moreover, a similar argument is part of Corollary \ref{cor:nakagamiHD}.
\end{rem} 
To incorporate Nakagami fading (cf. (\ref{eqn:naka_pdf})) we introduce for convenience a further notation, which is the link to \cite{Nakagami12}: For a finite index set $J\subset \mathcal N$ and an $\R$-valued tuple $\xi = [\xi_\nu]_{\nu\in J}$
 let


$$\begin{aligned}\gamma_{xy}^J(\xi) &= e^{-\theta\cdot m_{x} \cdot \mu_{xy} N} \sum_{s=0}^{m_{x}-1}\left( \theta \cdot m_{x} \cdot \mu_{xy} N \right)^s \\& \cdot \sum_{t=0}^{s}\frac{(\frac{q_{x}}{N})^t}{\left( s-t \right)!}   \cdot \sum_{\substack{\ell_\nu \ge 0\\ \sum\limits_{\nu \in J}\ell_\nu = t}}\prod_{\nu \in J}\left[ \vphantom{\frac{a^2}{b^2}}\left( 1-\xi_{\nu} \right)\delta_{0\ell_\nu} \right. \\
&\ds\ds +\left. \left(\begin{aligned} \ell_\nu + &m_\nu -1\\ &\ell_{\nu}\end{aligned}\right)\cdot \frac{\xi_{\nu}\cdot\left( \frac{1}{m_\nu q_{x} \mu_{\nu y}} \right)^{\ell_\nu}}{\left( \theta\frac{m_x}{m_\nu}\frac{\mu_{xy}}{\mu_{\nu y}}+1 \right)^{m_\nu+\ell_\nu} }  \right].
  \end{aligned}$$

\begin{cor}[Nakagami fading expectations and covariances]
\label{cor:nakagamiHD}
In the half-duplex model under Nakagami fading with pdf (\ref{eqn:naka_pdf}) the expectation for the link $u_i u_j,\, i \ne j,$ is given by
	$$\E[ Y_{ij}] = \left(1-\frac{1}{m}\right) \cdot \gamma_{ij}^{\mathcal N\setminus \{i,j\}}(\one/m).$$
For $i = k$ the covariance of the link $u_i u_j$ with link $u_k u_l$ is given by (\ref{eqn:cov_a_HD}) using (\ref{eqn:state_weight_a_HD}) and the state coefficients
	$$\begin{aligned}\alpha_{xyxw}(c) = \gamma_{xy}^{\mathcal N\setminus\{x,y,w\}}(c),
	\end{aligned}$$
	 	where $c = [c_\nu]_{\nu \in \mathcal{N}\setminus\{i,j,l\}}.$

For $i \ne k,$  it is given by (\ref{eqn:cov_b_HD}) using (\ref{eqn:state_weight_a_HD}), (\ref{eqn:state_weight_b_HD}) and the state coefficients
	$$\alpha_{xyzw}(c') = \gamma_{xy}^{\mathcal N\setminus \{x,y,w\}}(\xi), \ds c' = [c_\nu']_{\nu \in \mathcal{N}\setminus\{i,j,k,l\}},$$
	where $\xi = [\xi_\nu]_{\nu \in \mathcal N\setminus\{x,y,w\}},\,\xi_z =1,\, \xi_\nu = c'_\nu,\; \nu \ne z,$ and
	$$\beta_{xyzw}^{(d)}(c'')= \gamma_{xy}^{\mathcal N\setminus\{x,y,z\}}(\zeta), \ds c'' = [c_\nu'']_{\nu \in \mathcal{N}\setminus \{i,k\}},$$
 where  $\zeta = [\zeta_\nu]_{\nu \in \mathcal N\setminus\{x,y,z\}},\, \zeta_{\nu}  = \delta_{c''_\nu d}, d = 1,2.$
\end{cor}
\begin{proof}
This is a consequence of Theorem \ref{prp:slotCorrelationsHD} and   \cite{Nakagami12}. See 
\if\pubMode\IEEE
the technical report \cite{TechRep}
\else
Appendix  \ref{proof:cor_HD}
\fi
for details.
%
\end{proof}  

In Fig.~\ref{fig:correlation} we show an example plot resulting from Corollary~\ref{cor:nakagamiHD} with Nakagami fading shape parameter $2$ and $2$ slots.  
The plot shows the correlation matrix for all possible communication links among $6$ nodes arranged on a $2\times 3$ grid (cf. Section \ref{sec:parameters}). 


\color{black}

\color{black}
\section{Application to Average Consensus}
\label{sec:application}
\subsection{Discrete probabilistic average consensus}
In the discrete, linear, first order probabilistic average consensus protocol \cite{Olfati-Saber2007}, \cite{Fagnani08}, \cite{Silva11}, a group of $n$ nodes periodically exchanges broadcast messages each containing a sender's dataset. 
For simplicity we assume all nodes have scalar, real valued data. It is an iterative random process whose goal is that all nodes agree on the same value in the limit with probability one. Further this value, the \df{agreement value}, should be close to the average of the initial data. Nonetheless, due to asymmetries in packet loss between outgoing and incoming transmissions, the agreement value will differ from the true average
which, however, shall not be our concern here.
Instead we focus on a probabilistic description of all nodes' deviation from the \textit{current} average at \textit{every} iteration. We measure this deviation in terms of the \textit{root mean square} (RMS) error (or \df{disagreement}) after $k\in \N$ iterations (or steps) which we now introduce: 
If the (random) vector $x_k = (x_k^{(1)}, ..., x_k^{(n)}) \in \R^n$ represents the nodes' states after $k$ consensus iterations having started from the deterministic initial state $x_0\in \R^n,$ then by $\sum_{i = 1 }^{n} \left( x_k^{(i)} - \frac{1}{n} x_k^T \one \right)^2$ a scalar random variable will be given which we shall denote by $\delta_k^2(x_0).$
	Divided by $\sqrt{n-1},$ the entity $\delta_k(x_0)$ is just the \textit{sample standard deviation} of the components $x_k^{(i)}, i = 1,\dots,n,$ explaining why it is also called \df{disagreement}. See e.g.\cite{Fagnani08} for further reading.
For any initial state $x_0$ the RMS error after $k$ steps is defined as $\sqrt{\E \delta_k^2(x_0)}.$ 

Finally, the \df{maximum root mean square} (mRMS) error is the RMS error for the worst possible choice of inital state inside the closed unit ball, $\max_{\n{x} \le 1}\sqrt{\E \delta_k^2(x)}.$ Note that the RMS error attains its maximum for \df{normalized} initial values, i.e. those on the unit sphere.

\subsection{Performance measure}
In this subsection we use slightly modified results from the literature to bound the mRMS error from above\cite{ICCCN19}  and below\cite{Ogura2013}. 
The mRMS error stochastically describes the consensus disagreement at the $k$-th step for the worst possible, step dependent choice of the initial state. It  constitutes a performance measure capable of well capturing the outage correlation effects.
In \cite{ICCCN19} an uncorrelated half duplex Bernoulli model (UHBM) with symmetric packet losses was assumed. 
This assumption, however, does not affect the generality of the proof of the $k$-step estimates presented there, although some adaptions in notation are required.

To quantify the impact of outage correlation on consensus, let us introduce $n^2-n$ uncorrelated Bernoulli variables $Z_{ij}, i \ne j,$ rendering an asymmetric 
UHBM having the property\footnote{The symmetric UHBM of \cite{ICCCN19} has the alternative property $\E Z_{ij} = \E Z_{ji} = \E (Y_{ij} + Y_{ji})/2$ which we do not use here.}
	$\E [Z_{ij}] = \E [Y_{ij}],$
where $Y_{ij}$ are defined in section IV.
In particular $\cov(Z_{ij}, Z_{kl})=0,$ for $i \ne j, k \ne l$ and $(i,j) \ne (k,l).$

To apply the UHBM and our correlated half-duplex Bernoulli model (CHBM) in the consensus context, we need to link the uncorrelated variables $Z = (Z_{ij})_{i\ne j}$ and their correlated counterparts $Y = (Y_{ij})_{i\ne j}$ to so called \textit{Laplacian} valued random matrices $L^{Z}$ and $L^{Y}$ respectively: For any ensemble of Bernoulli variables $W = (W_{ij}),\, 1 \le i,j \le n, i \ne j,$ define $L^W= (\ell_{ij}^W)$ by 
	$$ \ell_{ij}^W = \left\{ \begin{aligned} &\sum_{\substack{k=1\\k\ne i}}^{n} W_{ki}, \ds &\mbox{ if } i = j,\\
		&-W_{ji}, \ds &\mbox{ if  } i \ne j  .\end{aligned}\right.$$
Note that for any $\varepsilon >0$ and any sequence $W^{(1)}, W^{(2)}, \dots$ of realizations\footnote{More precisely, $(W^{(1)}, W^{(2)},\ldots)$ is supposed to be a random sample with $W^{(k)}\stackrel{\text d}{=} W,\; k\in\N$, where $\stackrel{\text d}{=}$ means equality in distribution.}  of $W$ the discrete consensus protocol \cite{Olfati-Saber2007}, \cite{Silva11}, (also cf. \cite{ICCCN19}) is then given by the matrix iteration
	$$ x_{k+1} = (I_n - \varepsilon \cdot L^{W^{(k)}})\cdot  x_k,$$
	provided the values $W_{ij} = 0, 1$ are interpreted as unsuccessful/successful links $u_i u_j,$ i.e. $W$ represents the offdiagonal entries of a (transposed) adjacency matrix of an (in general undirected) graph (without loops); $x_0 \in\R^n$ is some arbitrary initial value. We call the number $\varepsilon$ \df{(discrete) gain}.
	We assume the realizations $W^{(1)}, W^{(2)}, \dots$ to be mutually independent, which reflects the IBF assumption. Note that from the communication perspective, this constitutes a sequence of frames; the slots play only an implicit role causing the correlation within any of the $W^{(k)}$'s.
Now let
	$$R^W(\varepsilon) :=  (\Pi \otimes \Pi) \cdot [(I_n- \varepsilon \E L^W)\otimes(I_n- \varepsilon \E L^W) + \varepsilon^2 C^W] $$
where the matrix $C^W\in \R^{n^2 \times n^2}$ is given in terms of its components $C^W_{IJ} = \cov(\ell_{ij}^W, \ell_{kl}^W),$ for $I = (n-1)\cdot i +k , J = (n-1)\cdot j + l, \ds i,j,k,l = 1 ,\dots, n.$
Since the covariance is a bilinear form, these components can be easily determined in terms of $W.$ E.g. for $W=Y$ and $i = j$ and $k\ne l$ we have $\cov(\ell_{ij}^Y, \ell_{kl}^Y) = -\sum_{\substack{\nu = 1\\ \nu \ne i}}^{n} \cov(Y_{\nu i}, Y_{lk}).$ Similarly, $\E [L^W]$ can be evaluated.
Finally, we can define the $L^2$-joint spectral \cite{Ogura2013} and numerical radius \cite{ICCCN19}, respectively ,
	$$r_2(W,\varepsilon) := \sqrt{\varrho(R^W(\varepsilon))} \mbox{ and } w_2(W, \varepsilon) := \sqrt{w(R^W(\varepsilon))},$$
where $\varrho(\,\cdot\,)$ denotes the spectral radius (i.e. the largest modulus of all eigenvalues) and $w(\,\cdot\,)$ denotes the numerical radius, e.g. \cite{He1997a}, of a matrix. 

The following theorem introduces an upper and a lower bound for the latter of which there always exists a normalized initial state $\hat x$ (i.e. $\n{\hat x} = 1)$ such that $\sqrt{\E \delta_k^2(\hat x)}$ exceeds or attains the lower bound. On the other hand, for \textit{any} normalized initial state $x$ 
the upper bound dominates $\sqrt{\E \delta_k^2(x)}.$

\begin{thm}[cf. \cite{ICCCN19} and e.g. \cite{Ogura2013}]\label{thm:bounds}
Let $W = Y,Z$ represent the 
CHBM or the UHBM.
Then the mRMS disagreement after $k$ iterations is bounded as follows:
$$\frac{r_2^k(W, \varepsilon)}{\sqrt n} \le \max_{\n{x} \le 1}\sqrt{\E \delta_k^2(x)} \le \sqrt{2\sqrt{n-1}}\cdot w_2^k(W, \varepsilon).$$
\end{thm} 
\begin{proof}
	The left-hand inequality follows from the infimum property of $r_2$, e.g.\cite[Lem. 2.7]{Ogura2013}), the right-hand one from \cite{ICCCN19} up to a factor of 
	$\sqrt[4]{(n-1)/n}.$ See 
	\if\pubMode\IEEE
    	\cite{TechRep}
    \else
    Appendix \ref{prf:thm:bounds}
    \fi
    for details.
\end{proof}


In analogy to the geometric mean we now define the following lower and upper performance bounds for the \df{average per step}  mRMS error, $\max_{\n{x} \le 1}\sqrt[2k]{\E \delta_k^2(x)}:$ 
\begin{equation}\label{eqn:bounds}\text{lb}(\varepsilon)= \frac{r_2(W,\varepsilon)}{\sqrt[2k]{n}}\; \mbox{ and }\;
\text{ub}(\varepsilon) = \sqrt[2k]{2 \sqrt{n-1}} \cdot w_2(W,\varepsilon),\end{equation}
where $\text{lb}(\varepsilon) =\text{lb}(\varepsilon;k,W)$ and $\text{lb}(\varepsilon) = \text{lb}(\varepsilon; k,W).$ 
\begin{rem}
\label{rem:thm_bounds}
\begin{rom}
\item
Asymptotically, the lower bound is tight since $r_2(W,\varepsilon) = \lim\limits_{k\sra \infty}\max_{\n{x} \le 1}\sqrt[2k]{\E \delta_k^2(x)},$
\if\pubMode\IEEE
cf.\cite{TechRep}.
\else
(cf. Appendix \ref{sec:appendix_additional}).
\fi
\item The number of steps $k$ required to obtain a prescribed precision
increases only logarithmically with the number of nodes $n$ (apart from the spread of $r_2$ and $w_2$). 

\item The upper bound used in \cite{Silva11} is simply the squared mRMS disagreement for the first step ($k=1$),  \if\pubMode\IEEE
cf. \cite{TechRep}.
\else
cf. Appendix \ref{sec:appendix_additional}.
\fi
\end{rom}
\end{rem}


\subsection{Exemplary effects for a selected deployment and parameters}
\label{sec:parameters}
We adopt network parameters similar to the IEEE 802.11p standard: A carrier frequency of 5.9 GHz and a bandwidth $B$ of 10 MHz. We choose a reference bitrate $R_0$ of 40\% of the 802.11p maximum bitrate of 27 Mbit/s. To obtain comparable consensus performance results, we let the beacon time interval be constant in terms of slots, more precisely we set $R$ according to the number of slots $m$ in use: $R = m \cdot R_0.$ The threshold $\theta$ is then computed according to Shannon capacity formula: $\theta = 2^{R/B}-1.$ For $m=1$ we obtain $\theta = 2.23$ such that the requirement $\theta \ge 1$ is always fulfilled. The thermal noise parameter $N$ is obtained from the Boltzmann constant $k_B$ at a reference temperature $T$ of 293K: $N = B\cdot k_B \cdot T.$

Our nodes are deployed on a regular $2\times 3$ grid having a horizontal and vertical distance of 1500m between neighbors, a common transmit power of 500 mW, a common varying Nakagami shape paramter $m_i = m_S, i = 1 , \ldots, n,$ and a path loss coefficient of 2.0.

\begin{figure}[t!]
	
	\centering
	\subfloat[\label{fig:performance_bounds_by_slots}Performance for different frame lengths $m.$ ]{\includegraphics[trim=080 5 80 40,clip, width = 0.49 \textwidth]{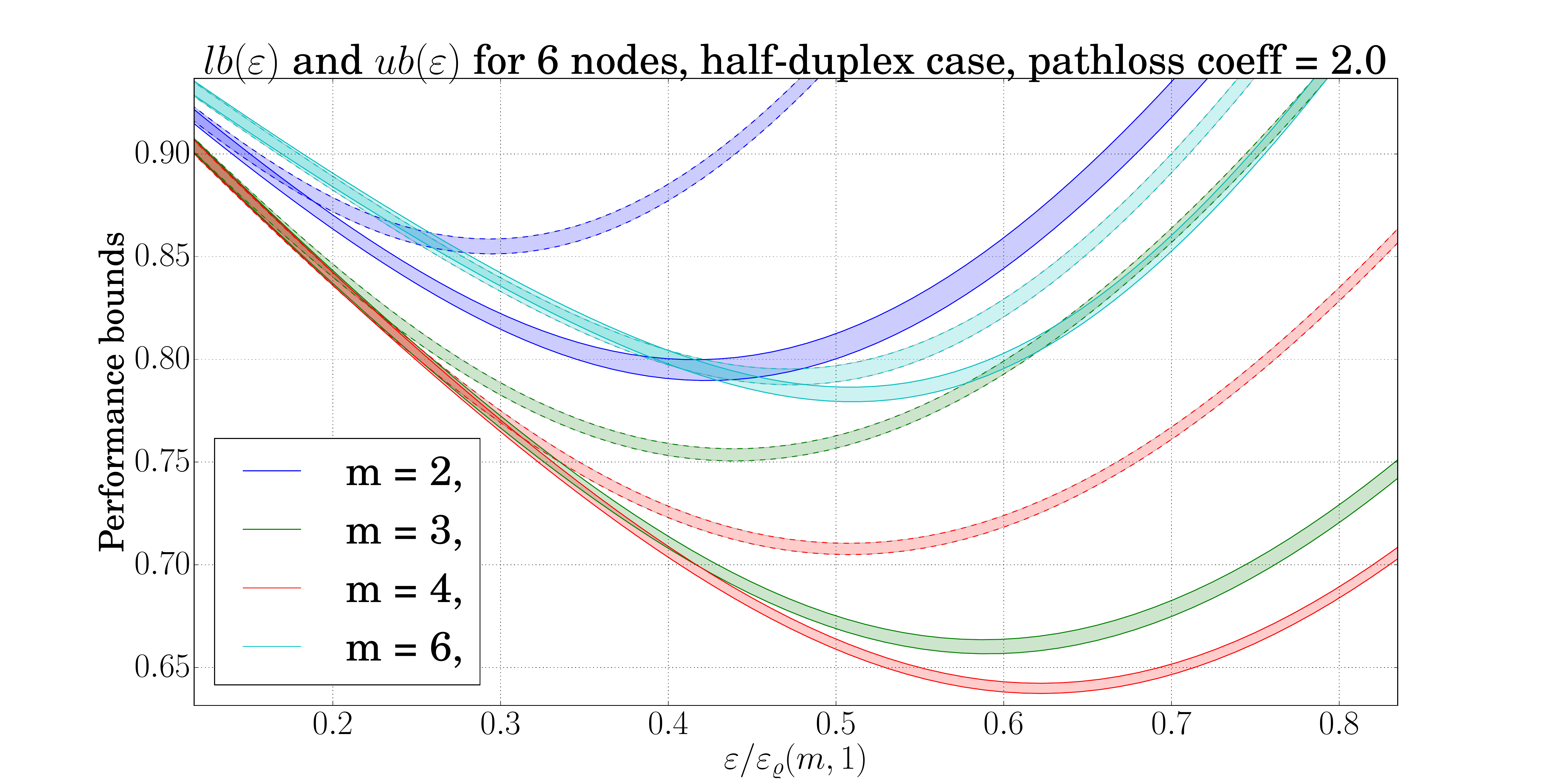}}\hfill
	\subfloat[\label{fig:performance_bounds_by_shape}Performance for different Nakagami shape parameters $m_S.$]{\includegraphics[trim=080 5 80 40,clip, width = 0.49 \textwidth]{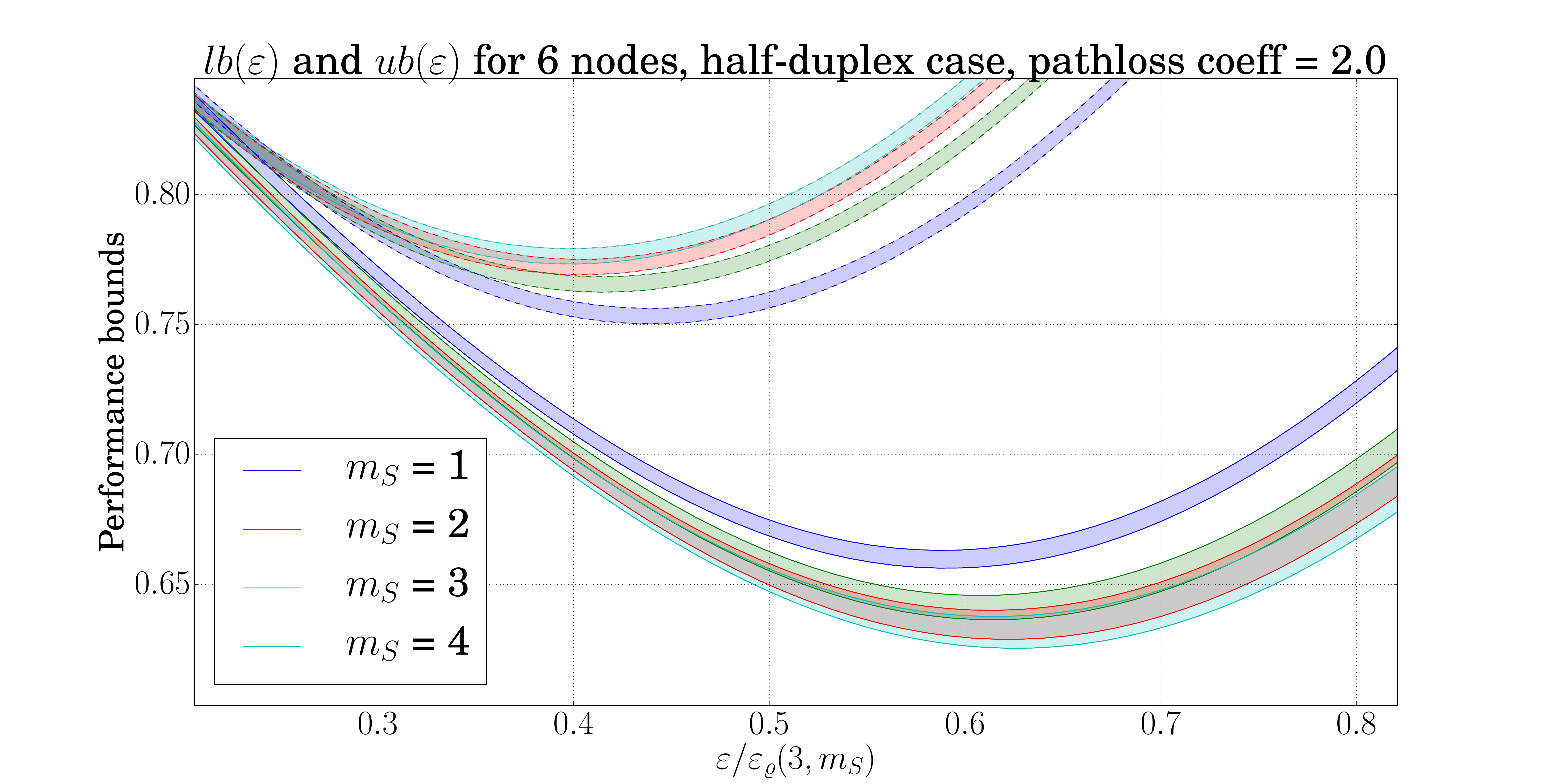}}\hfill 

\caption{\scriptsize Comparison of the consensus performance for uncorrelated ($W=Z,$ dashed curves) and correlated model ($W=Y,$ solid curves) in terms of the upper and lower bounds (\ref{eqn:bounds}) for $k = \numberOfSteps$ steps. On the abscissa we lay off the gain parameter scaled by the optimal essential spectral radius for the pair $(m,m_S),$ on the ordinate the dimensionless bounds (\ref{eqn:bounds}). Smaller values correspond to better performance, a value below (above) 1 for \textit{both} bounds implies (in-)stability in the exponential mean square sense \cite{Ogura2013}. Further, at a scope of $k$ steps the average per step mRMS error
lies within the colored regions. In Figure \ref{fig:performance_bounds_by_slots} the shape parameter $m_S$ is set to $1$ (Rayleigh fading) while the number of slots per frame $m$ varies. In Figure \ref{fig:performance_bounds_by_shape} the frame length $m$ is fixed to $3$ while the shape parameter $m_S$ varies.}
\end{figure}

Figures~\ref{fig:performance_bounds_by_slots} and \ref{fig:performance_bounds_by_shape} show
the consensus performance (\ref{eqn:bounds}) over the gain parameter which is scaled in terms of the \df{minimizer} $\varepsilon$ of the \df{essential spectral radius} $\varrho(\Pi- \varepsilon \E [L^W]),$  e.g. \cite{ICCCN19}.
The minimizer, denoted by $\varepsilon_\varrho(m,m_S),$ depends on the number of slots and the shape parameter unitilized, but is the same for the correlated and uncorrelated model, $W = Y,Z.$ 
We set $k$ to \numberOfSteps\ steps.

In Figure~\ref{fig:performance_bounds_by_slots} the number of slots is varied, in Figure~\ref{fig:performance_bounds_by_shape} the shape parameter.
We observe two effects. (i) In both figures the correlated model shows significantly better performance and also different locations of the minimizing gain setting $\varepsilon$ (the minimizer of the correlated model is closer to $\varepsilon_\varrho$). (ii) Remarkably, in Figure~\ref{fig:performance_bounds_by_shape} the order of the curves in terms of performance is inverted: Curves
 with lower shape parameter perform better in the uncorrelated model while they perform worse in the correlated one. 
While we observed effect (i) for various parameter settings (Figure~\ref{fig:performance_bounds_by_slots}, $m=6$ constitutes a borderline case), effect (ii) is rather unusual.



\section{Conclusion}\label{sec:conclusion}
We have derived closed form expressions for the outage/success correlation in a framed slotted ALOHA scheme under the Nakagami fast fading model for fixed relative node positions.
The found expressions enable a thorough performance analysis of distributed control applications in wireless networks for which we provide bounds in an appropriate form.

Although inaccurate compared to the correlated model, the uncorrelated one 
could still serve 
as rough but simpler structured bound for the control performance. How far it is sufficient to use the simpler bound requires a broader study.

Our method can in principle be carried over to related slotted ALOHA protocols, e.g. spatial and opportunistic spatial ALOHA \cite{Baccelli2009a}. 
Apart from the broadcast case, an application in interference cancellation\cite{Clazzer2017} is also thinkable.



\bibliography{bibfile}{}
\bibliographystyle{plain}

\if\pubMode\TechRep

\newpage

\appendix
Here we provide the proofs using our slot framework. We start simple by proving a remark, moving on to the full duplex case until we finally arrive at the more complex situation in the half duplex case. The remainder of the Appendix deals with the details of the consensus performance bounds, in particular the proof of Theorem \ref{thm:bounds}.

In the following, let $(\Omega, \mathcal{F}, \Pr)$ be the underlying probability space.

\subsection{Proof of Remark \ref{rem:backward_compatibility}}
\label{prf:backward_compatibility}
Let us begin with another general remark about conditioning on independent slot variables $S_i, S_j.$ Let $A\in \mathcal{F}$ be an arbitrary event. Then $\Omega$ can be disjointly partitioned as
    $$\Omega = \dot \bigcup_{b_i,b_j = 1}^m \{S_i = b_i, S_j = b_j\}$$
and we have, using $\Pr(S_i\ne S_j) = \frac{m-1}{m}$ and $\Pr(S_i=b_i; S_j =b_j) = \frac{1}{m^2},$
    \begin{equation}\label{eqn:remark_independent_slots} \begin{aligned}&\Pr(A \; \mid \; S_i \ne S_j)\\
    &\ds= \frac{\Pr(A \mbox{ and } S_i \ne S_j)}{\Pr(S_i \ne S_j)}\\
    &\ds= \frac{m}{m-1}\sum\limits_{b_i, b_j=1}^{m}\Pr(A \mbox{ and } (S_i \ne S_j; S_i = b_i; S_j = b_j))\\
    &\;\;= \frac{m}{m-1}\sum\limits_{b_i \ne b_j}\Pr(A \mid S_i =b_i; S_j=b_j) \cdot \Pr(S_i = b_i; S_j = b_j)\\
    &\ds= \frac{1}{m(m-1)}\sum\limits_{\substack{b_i, b_j=1\\b_i\ne b_j}}^{m}\Pr(A \mid S_i =b_i;\; S_j=b_j).
    \end{aligned}
\end{equation}

Using the law of total probability and (\ref{eqn:remark_independent_slots}) we now establish the connection between full duplex and half duplex success probabilities. For $i \ne j$ we have from (\ref{eqn:X_Y})
\begin{equation}\label{eqn:expectation_decomp} \begin{aligned}\E [Y_{ij}] &=
		\frac{1}{m} \cdot 0 + \left( 1-\frac{1}{m} \right) \E\left[
		X_{ij} \mid S_i \ne S_j \right]\\
	& = \frac{1-\frac{1}{m}}{m(m-1)}\sum_{\substack{b_i, b_j =1\\b_i \ne b_j}}^{m}\E \left[ X_{ij}^{S_i} \mid S_i = b_i;\, S_j = b_j \right] \\
	& = \frac{1-\frac{1}{m}}{m(m-1)}\sum_{\substack{b_i, b_j =1\\b_i \ne b_j}}^{m}\Pr \left(\frac{P_{ij}}{N + \sum_{\nu \ne i,j}^{}\delta_{b_i S_\nu}P_{\nu j}}\ge \theta  \right) \\
	& = \frac{1-\frac{1}{m}}{m}\sum_{b_i =1}^{m}\Pr \left(\frac{P_{ij}}{N + \sum_{\nu \ne i,j}^{}\delta_{b_i S_\nu}P_{\nu j}}\ge \theta  \right) \\
	& = \left(1-\frac{1}{m}\right)\cdot \Pr \left(\frac{P_{ij}}{N + \sum_{\nu \ne i,j}^{}\delta_{S_i S_\nu}P_{\nu j}}\ge \theta  \right) \\
	& = \left( 1-\frac{1}{m} \right) \cdot \E [X_{ij}]
	.\end{aligned}
\end{equation}
To verify that in the Rayleigh fading case\footnote{The consistence with \cite{Nakagami12} in the general case is shown in Corollary \ref{cor:nakagamiFD}.} $\E X_{ij}$ is consistent with the expressions derived in \cite{Nakagami12} and \cite{ICCCN19}, we assume that $X_{ij}$ is distributed according to the pdf (\ref{eqn:naka_pdf}) with shape parameter $m_i = 1$ for all $i = 1, \dots n.$ 
Utilizing a version of the conditioned expectation for any nonnegative, integrable random variable $Q$ independent of $P_{ij}$ yields 
	\begin{equation}\label{eqn:conditioned_expectation}\begin{aligned}&\Pr\left(\frac{P_{ij}}{N+ Q} \ge \theta\right) \\
		&= \E [\;\E\left[ \ind\left( P_{ij} > \theta( N + Q) \right) \mid Q \right]]  \\
		&= \E [\exp\left\{ -\mu_{ij} \theta(N + Q) \right\}],
	\end{aligned}
	\end{equation}
where we have used 
	$$\E [\ind\left( P_{ij} > \theta(N+x) \right)] = e^{-\mu_{ij}\theta(N+x)},\; x \ge 0.$$
 Thus, for $Q = \sum_{l\ne i,j}\delta_{S_l S_i} P_{l j},$ we get

 	$$\begin{aligned} \E [X_{ij}] &= \E [X_{ij}^{S_i}] = \Pr\left(\frac{P_{ij}}{N+\sum_{l\ne i,j}\delta_{S_l S_i} P_{l j}} \ge \theta\right)\\
		&\stackrel{\text{(a)}}{=} e^{-\mu_{ij}\theta N }\cdot \E \left[\prod_{l\ne i,j}\exp\left\{ -\mu_{i j} \theta\delta_{S_i S_l} P_{l j}  \right\}\right]\\
		&\stackrel{\text{(b)}}{=} \frac{e^{-\mu_{ij}\theta N }}{m}\cdot\sum_{b_i = 1}^{m} \E\left[ \prod_{l\ne i,j}\exp\left\{ -\mu_{i j} \theta\delta_{b_i S_l} P_{l j}  \right\}\right]\\
		&\stackrel{\text{(c)}}{=} \frac{e^{-\mu_{ij}\theta N }}{m}\cdot\sum_{b_i = 1}^{m} \prod_{l\ne i,j}\E[\exp\left\{ -\mu_{i j} \theta\delta_{b_i S_l} P_{l j}  \right\}]\\
		&= \frac{1}{m} e^{-\mu_{i j} \theta N }\sum_{b_i=1}^{m}\prod_{l\ne i,j}  \frac{1}{m}\sum_{b_l = 1}^{m}\E [\exp\left\{ -\mu_{ij } \theta \delta_{b_i b_l} P_{l j} \right\}]\\
		&= \frac{1}{m} e^{-\mu_{i j} \theta N } \sum_{b_i=1}^{m}\prod_{l\ne i,j} \left( \frac{m - 1}{m} + \frac{1}{m}\E [e^{-\mu_{i j}\theta P_{lj}}] \right)\\
		&\stackrel{\text{(d)}}{=}  e^{-\mu_{i j} \theta N } \prod_{l\ne i,j} \left( (1-\frac{1}{m}) + \frac{1}{m}\cdot\frac{1}{1+\theta \frac{\mu_{i j}}{\mu_{lj}}} \right)\\
		&=  e^{-\mu_{i j} \theta N } \prod_{l\ne i,j} \left( 1 - \frac{1}{m}\cdot \frac{\theta \frac{\mu_{i j}}{\mu_{lj}}}{\theta \frac{\mu_{i j}}{\mu_{lj}} +1} \right)\\
		&=  e^{-\mu_{i j} \theta N } \prod_{l\ne i,j} \left( 1 - \frac{1}{m} \cdot \frac{\theta}{\theta + \frac{\mu_{lj}}{ \mu_{i j} }} \right),
	\end{aligned}$$ 
where we have used (\ref{eqn:conditioned_expectation}) in (a), the independence of the slot variables combined with the IBF assumption in (b), (c) and the identity
	$$ \E [\exp\left\{ -\mu_{ij}\theta P_{lj} \right\}]  = \frac{1}{1+\theta \frac{\mu_{ij}}{\mu_{lj}}},$$
in (d) which is specific for the Rayleigh distribution. Hence, together with (\ref{eqn:expectation_decomp}) we obtain
$$ \E [Y_{ij}] =  \left( 1-\frac{1}{m} \right) \cdot e^{-\mu_{i j} \theta N } \prod_{l\ne i,j} \left( 1 - \frac{1}{m} \cdot \frac{\theta}{\theta + \frac{\mu_{lj}}{ \mu_{i j} }} \right).$$

	\qed

\subsection{Full duplex variant}
\label{sec:full_duplex}
We now establish the simpler full duplex analogs of Theorem \ref{prp:slotCorrelationsHD} and Corollary \ref{cor:nakagamiHD} of section \ref{sec:corr} since the idea of the method becomes clearer in this case.
\begin{thm}[Full-duplex covariances]
\label{prp:slotCorrelationsFD}
Let $\theta \ge 1,N > 0, i \ne j, k \ne l,$ and $(i,j) \ne (k,l).$ 

\begin{abc}\item We have 
		$$\E[X_{ij}X_{kl} \vert S_i = S_k] = \frac{1-\delta_{lj}}{m^{n-2+\delta_{ik}}} \cdot \Phi_{ijkl}^{\mathcal N \setminus\left\{ i,k \right\}}(\phi, \alpha).$$
	where the state weight $\phi(c)$ is given by
		\begin{equation}\label{eqn:state_weight_a_FD} \phi(c) = \prod_{\nu \in \mathcal N \setminus \left\{ i,k \right\}}(m-1)^{1-c_\nu},\end{equation}
	the state coefficients $\alpha_{xyzw}(c) =\; \alpha_{xyz}(c)$ are given by 
	\begin{equation*}
		\begin{aligned}
		\Pr\left(\frac{P_{xy}}{N+\sum\limits_{\nu\ne x,y,z}^{}c_\nu P_{\nu y} + (1-\delta_{xz})P_{zy}} \ge \theta\right),		\end{aligned}
	\end{equation*}
	and where $c = [c_\nu]_{\nu \in \mathcal N \setminus \left\{ i,k \right\}}.$

	In particular, $\E [X_{ij}X_{il}] = \E [X_{ij}X_{il} \vert S_i = S_i]$ can be expressed this way which gives for $j \ne l$
	\begin{equation}\label{eqn:cov_a_FD}\cov(X_{ij}, X_{il}) = \frac{ \Phi_{ijil}^{\mathcal N\setminus \left\{ i \right\}}(\phi, \alpha)}{m^{n-2}}  - \E [X_{ij}] \E [X_{il}].\end{equation}
		Particularly, $c = [c_\nu]_{\nu \in \mathcal N\setminus\left\{ i \right\}} \in \left\{0,1  \right\}^{n-1}.$

  	\item For $k\ne i,$
		$$ \begin{aligned}&\E [X_{ij}X_{kl}\vert S_i \ne S_k]  = \frac{1}{m^{n-2}}\cdot \Psi_{ijkl}
			,\end{aligned}$$
		hence
		\begin{equation}\label{eqn:cov_b_FD} \begin{aligned} \cov(X_{ij}, X_{kl}) &= \frac{1-\delta_{lj}}{m^{n}} \cdot \Phi_{ijkl}^{\mathcal N \setminus \left\{ i,k \right\}}(\phi, \alpha)\\
			&\;\,    + \frac{m-1}{ m^{n-1}}\cdot \Psi_{ijkl}^{\mathcal N \setminus\left\{ i,k \right\}}(\psi, \beta) - \E[X_{ij}] \E [X_{kl}],  \end{aligned}\end{equation}
	    where $\Phi_{ijkl}(\phi, \alpha)$ depends on the state functions of part (a) whence $\Psi_{ijkl}(\psi, \beta)$ depends on the state weight
	    	\begin{equation}\label{eqn:state_weight_b_FD}\psi(c) = \prod_{\substack{\nu = 1\\ \nu \ne i,k}}^n (m-2)^{\delta_{c_\nu, 0}}\end{equation}
	and on the state coefficients
		$$\beta_{xyzw}^{(d)}(c)  = \Pr\left(\frac{P_{xy}}{N+\sum_{\nu\ne x,y,z}^{}\delta_{c_\nu d} P_{\nu j'}} \ge \theta  \right),$$
where $d = 1,2.$
 
	\end{abc}
\end{thm} 
\begin{proof}

We will make frequent use of (\ref{eqn:remark_independent_slots}) and related rearrangements.
    
\begin{abc}
	\item Let $i \ne j, k\ne l.$ Assume first $i \ne k.$ Then
		$$\begin{aligned} &\E \left[ X_{ij} X_{kl} \mid S_i = S_k \right] \\
			&= \Pr\left( \frac{P_{ij}}{N + \sum_{\substack{\nu= 1\\\nu\ne i,j}}^{n}\delta_{S_i S_\nu} \cdot P_{\nu j}} \ge \theta \mbox{ and } \right.\\
			&\left.\left. \ds \ds \frac{P_{kl}}{N + \sum_{\substack{\nu= 1\\\nu\ne k,l}}^{n}\delta_{S_k S_\nu} \cdot P_{\nu l}} \ge \theta  \right\vert S_i = S_k \right)\\
			&= \frac{1}{m^{n-1}} \sum_{\substack{b_i = 1\\ (b_i = b_k)}}^{m}\sum_{\substack{b_\nu = 1\\ \nu \ne i,k}}^{m} \left[ \vphantom{\frac{A}{B}}\right.\\
			& \; \cdot \Pr\left( \frac{P_{ij}}{N + \sum_{\substack{\nu= 1\\\nu\ne i,j,k}}^{n}\delta_{b_i b_\nu} \cdot P_{\nu j} + P_{kj}} \ge \theta \mbox{ and } \right.\\ 
			&\ds \ds \left.\left. \ds \ds \frac{P_{kl}}{N + \sum_{\substack{\nu= 1\\\nu\ne k,l,i}}^{n}\delta_{b_i b_\nu} \cdot P_{\nu l} + P_{il}} \ge \theta  \right)\right].\\
		\end{aligned}$$
	In case $j = l,$ observe that the summands in the previous step reflect the following situation, with $a,b > 0$ appropriately set:
		$$\Pr\left( \frac{a}{N+b} \ge \theta \mbox{ and }  \frac{b}{N + a} \ge \theta  \right) = \Pr(\emptyset) =  0,$$
	which holds since $N > 0,\; \frac{a}{N+b} \ge \theta \ge 1 \mbox{ and }  \frac{b}{N + a} \ge \theta \ge 1$ leads to the contradiction
		$$a>b \mbox{ and } b > a.$$
Hence we have to add a prefactor $1- \delta_{jl}$ and we can focus on the situation when $j\ne l.$ In this case the summands split into two probability factors since the powers received at the different nodes $u_j$ and $u_l$ are independent due to the IBF assumption.
Together, this gives
		$$\begin{aligned} &\E \left[ X_{ij} X_{kl} \mid S_i = S_k \right] \\
			&= \frac{1- \delta_{jl}}{m^{n-1}} \sum_{\substack{b_i = 1\\ (b_i = b_k)}}^{m}\sum_{\substack{b_\nu = 1\\ \nu \ne i,k}}^{m}\left[\vphantom{\frac{A}{B}}\right. \\
			&\; \cdot \Pr\left( \frac{P_{ij}}{N + \sum_{\substack{\nu= 1\\\nu\ne i,j,k}}^{n}\delta_{b_i b_\nu} \cdot P_{\nu j} + P_{kj}} \ge \theta \right)\\ 
			&\; \cdot \left.\Pr\left(\frac{P_{kl}}{N + \sum_{\substack{\nu= 1\\\nu\ne k,l,i}}^{n}\delta_{b_i b_\nu} \cdot P_{\nu l} + P_{il}} \ge \theta  \right)\right].\\
		\end{aligned}$$ 
Inspection of the summands reveals that the $m^{n-1}$ states of the $b_\nu$'s$, \nu \ne k$ can be cast into $n-2$ macro state variables $c_\nu = 0, 1,\; \nu \in \left\{ 1, \dots, n  \right\}\setminus \{i,k\} $ representing the cases $b_i \ne b_\nu$ and $b_i = b_\nu$ respectively. For each $b_\nu$ the former case has $n-1$ realizations while for the latter there is only one. Thus, the combined weight for the macro state $c = (c_{\nu}), \nu \ne i,k,$ is 
		$$\phi(c) = \prod_{\nu \ne i,k}(m-1)^{1-c_\nu}.$$
We can now express $\E \left[ X_{ij} X_{kl} \mid S_i = S_k \right]$ in terms of the macro states:
$$ \begin{aligned}&\E \left[ X_{ij} X_{kl} \mid S_i = S_k \right] \\
	& = \frac{1- \delta_{jl}}{m^{n-1}} \sum_{\substack{b_i = 1\\ (b_i = b_k)}}^{m}\sum_{\substack{c_\nu = 0\\ \nu \ne i,k}}^{1}\left[\vphantom{\frac{A}{B}}\right. \\
	& \; \cdot \Pr\left( \frac{P_{ij}}{N + \sum_{\substack{\nu= 1\\\nu\ne i,j,k}}^{n}c_{\nu} \cdot P_{\nu j} + P_{kj}} \ge \theta \right)\\ 
	& \; \cdot \left.\Pr\left(\frac{P_{kl}}{N + \sum_{\substack{\nu= 1\\\nu\ne k,l,i}}^{n}c_{\nu} \cdot P_{\nu l} + P_{il}} \ge \theta  \right)\right]\\
 	& =  \frac{1- \delta_{jl}}{m^{n-2}} \sum_{\substack{c_\nu=0\\\nu\ne i,k }}^{1}\phi(c)\cdot \alpha_{ijk}(c)\cdot \alpha_{kli}(c). \end{aligned}$$
In case $i = k$ the summands $P_{kj}$ and $P_{il}$ in the denominator of the factors of the product of probabilities vanish which has already been incorporated in the $\alpha_{xyz}(x)$'s in virtue of the Kronecker delta. Moreover, since the slot variables $S_i$ and $S_k$ now coincide, the number of possible slot states is fewer by a factor of $m$ which leads to another Kronecker delta $\delta_{ik},$ placed in the denominator of the sum's prefactor.
 
	\item We now treat the case $S_i \ne S_k,$ which implies $i \ne k.$ Let $i\ne k.$ Then we have
		\begin{equation}\label{eqn:case_b_independence}\begin{aligned}&\E\left[ X_{ij}X_{kl}\vert S_i \ne S_k \right]\\
			&= \Pr\left( \frac{P_{ij}}{N + \sum_{\substack{\nu= 1\\\nu\ne i,j}}^{n}\delta_{S_i S_\nu} \cdot P_{\nu j}} \ge \theta \mbox{ and } \right.\\
			&\left.\left. \ds \ds \frac{P_{kl}}{N + \sum_{\substack{\nu= 1\\\nu\ne k,l}}^{n}\delta_{S_k S_\nu} \cdot P_{\nu l}} \ge \theta  \right\vert S_i \ne S_k \right)\\
		&= \frac{1}{m^{n-1}(m-1)} \sum_{\substack{b_i, b_k = 1\\ b_i \ne b_k}}^{m}\sum_{\substack{b_\nu = 1\\ \nu \ne i,k}}^{m} \left[\vphantom{\frac{A}{B}}\right.\; \\
		& \ds \ds \cdot  \Pr\left( \frac{P_{ij}}{N + \sum_{\substack{\nu= 1\\\nu\ne i,j,k}}^{n}\delta_{b_i b_\nu} \cdot P_{\nu j}} \ge \theta \right)\\
		&\ds \ds \cdot\left. \Pr\left( \frac{P_{kl}}{N + \sum_{\substack{\nu= 1\\\nu\ne k,l,i}}^{n}\delta_{b_k b_\nu} \cdot P_{\nu l}} \ge \theta  \right)\right]
		\end{aligned}\end{equation}
	Where we have exploited that the interference terms in both fractions are disjoint and so the fractions are independent.
	
	Now we introduce macro states $c_{\nu}:$ The state $c_\nu = 1$ represents the case $b_\nu = b_i$, the state $c_\nu = 2$ represents the case $b_\nu = b_k,$ and the state $c_\nu = 0$ represents any other case. Hence the weights for $c_\nu = 0, 1, 2$ are $m-2, 1, 1$ respectively which are gathered in the factor $\psi(c).$ Continuing from (\ref{eqn:case_b_independence}) we therefore obtain
		$$\begin{aligned}
		    &\E\left[ X_{ij}X_{kl}\vert S_i \ne S_k \right]\\
			&= \frac{1}{m^{n-1}(m-1)} \sum_{\substack{b_i, b_k = 1\\ b_i \ne b_k}}^{m}\sum_{\substack{c_\nu = 0\\ \nu \ne i,k}}^{2} \psi(c)\; \cdot\left[\vphantom{\frac{A}{B}}\right. \\
			& \ds \ds \cdot \Pr\left( \frac{P_{ij}}{N + \sum_{\substack{\nu= 1\\\nu\ne i,j,k}}^{n}\delta_{c_\nu 1} \cdot P_{\nu j}} \ge \theta \right) \\
			&\ds \ds \cdot \left.\Pr\left( \frac{P_{kl}}{N + \sum_{\substack{\nu= 1\\\nu\ne k,l,i}}^{n}\delta_{c_\nu 2} \cdot P_{\nu l}} \ge \theta  \right)\right]\\
			&= \frac{1}{m^{n-1}(m-1)} \sum_{\substack{b_i, b_k = 1\\ b_i \ne b_k}}^{m}\sum_{\substack{c_\nu = 0\\ \nu \ne i,k}}^{2} \psi(c) \cdot \beta_{ijk}^{(1)}(c) \cdot \beta_{kli}^{(2)}(c)\\
			&= \frac{1}{m^{n-2}} \sum_{\substack{c_\nu = 0\\ \nu \ne i,k}}^{2} \psi(c) \cdot \beta_{ijk}^{(1)}(c) \cdot \beta_{kli}^{(2)}(c),
		\end{aligned}$$ 
	where in the last step we have used that all $m\cdot(m-1)$ summands of the first sum are equal.
\end{abc}

\end{proof}

\begin{cor}[Nakagami fading expectations and covariances, full-duplex case]
\label{cor:nakagamiFD}
In the full-duplex model under Nakagami fading with pdf (\ref{eqn:naka_pdf}) the expectation for the link $u_i u_j,\, i \ne j,$ is given by
	$$\E X_{ij} = \gamma_{ij}^{\mathcal N\setminus\{x,y\}}(\one/m).$$ 
For $i = k$ the covariance of the link $u_i u_j$ with link $u_k u_l$ is given by (\ref{eqn:cov_a_FD}) using (\ref{eqn:state_weight_a_FD}) and the state coefficients
$$\begin{aligned}\alpha_{xyx}(c) = \gamma_{xy}^{\mathcal N\setminus\{x,y\}}([c_\nu]_{\nu\in \mathcal N\setminus\{x,y\}}), \end{aligned}$$
	where $c = \left[ c_\nu \right]_{\nu \in \mathcal{N}\setminus \left\{ i \right\}}.$

For $i \ne k,$  it is given by (\ref{eqn:cov_b_FD}) using (\ref{eqn:state_weight_a_FD}), (\ref{eqn:state_weight_b_FD}) and the state coefficients
$$\alpha_{xyz}(c') = \gamma_{xy}^{\mathcal N\setminus\left\{ x,y \right\}}(\xi), \ds  c' = \left[ c'_\nu \right]_{\nu \in \mathcal{N}\setminus \left\{ i,k \right\}},$$
where $\xi = \left[ \xi_\nu \right]_{\nu \in \mathcal N\setminus \left\{ x,y \right\}},\, \xi_{\nu}  = c'_\nu$ if $\nu \ne z,\, \xi_{z} = 1,$ and
	$$\beta_{xyz}^{(d)}(c'') = \gamma_{xy}^{\mathcal N\setminus\left\{ x,y,z \right\}}(\zeta), \ds c'' = \left[ c''_\nu \right]_{\nu \in \mathcal{N}\setminus \left\{ i,k \right\}},$$
where $\zeta = \left[ \zeta_\nu \right]_{\nu \in \mathcal N\setminus \left\{ x,y,z \right\}},\, \zeta_{\nu}  = \delta_{c''_\nu d}.$
\end{cor}
\begin{proof}
The proof of the half duplex counterpart of this corollary runs along the same lines as this one. We only proof the former since it part of the main paper, cf. \ref{proof:cor_HD}.
\end{proof}

\subsection{Proof of Theorem \ref{prp:slotCorrelationsHD}}
\label{prf:slotCorrelationsHD}
In the following we use similar arguments as in the proof of Theorem \ref{prp:slotCorrelationsFD}. However, the condition 
\begin{equation}\label{eqn:condition_asterisk} \ds  S_i \ne S_j \mbox{ and } S_k \ne S_l \end{equation}
introduces greater complexity here. Let us thus first gather all cases in Table I. 
For once, in this table different node subscripts shall refer to different nodes, e.g. the situation that $u_i = u_j$ is not allowed in the table. Consider two links, \textit{link 1}  and \textit{link 2}, and denote by $J$ the set of all node indices belonging to link 1 or link 2. The check marks in Table I 
indicate that - under condition (\ref{eqn:condition_asterisk}) - the condition in the column's head is automatically fulfilled, the symbol $\myxmark$ indicates an impossible condition, and \textit{depends} means that whether the condition is met or not depends on the actual state of the slot variables $S_i$ and $S_k.$

From Table I we see that $\Pr(S_i \ne S_j \text{ and } S_k \ne S_l)$ equals $\left(1-\frac{1}{m}\right)^2$ if $(k,l) \ne (j,i)$ and $1-\frac{1}{m}$ otherwise. Thus
	\begin{equation}\label{eqn:tot_prob} \begin{aligned}\E [Y_{ij} Y_{kl}] = \left(1-\frac{1}{m}\right)^{2-\delta_{kj}\delta_{li}} \cdot \E [X_{ij} X_{kl} \mid S_i \ne S_j; S_k \ne S_l]. \end{aligned}\end{equation}

\begin{table}
\label{table:conditions}
\begin{center}
\begin{tabular}{c|c|c|c|c|c}
	link 1 &link 2 & $\abs{J}$ & condition (\ref{eqn:condition_asterisk}) &  $S_i = S_k$   &  $S_i \ne S_k$   \\ \hline
\hline
$u_i u_j$ &$u_iu_l$ & 3 & $S_i \ne S_j,S_l$ & $\checkmark$ & \myxmark \\
$u_i u_j$ &$u_ju_i$ & 2 & $ S_i \ne S_j $ & $\myxmark$ & \checkmark\\
$u_i u_j$ &$u_ju_l$ & 3 & $ S_j \ne S_i,S_l $ &$\myxmark$ & \checkmark\\
$u_i u_j$ &$u_ku_i$ & 3 & $ S_i \ne S_j,S_k $ &$\myxmark$ & \checkmark\\
$u_i u_j$ &$u_ku_j$ & 3 & $ S_j \ne S_i,S_k $ & depends &  depends \\
$u_i u_j$ &$u_ku_l$ & 4 & $ S_i\ne S_j;\; S_k \ne S_l $ & depends & depends\\ \hline
\end{tabular}
\end{center} 
\caption{\small Situation for different choices of links.}
\end{table}

\begin{table}
\label{table:slot_probabilites}
\begin{center}
\begin{tabular}{c|c|c|c}
	link 1 &link 2 & $\Pr(S_i = S_k \,\mbox{ and }(\ref{eqn:condition_asterisk}) )$ & $\Pr(S_i \ne S_k \,\mbox{ and }(\ref{eqn:condition_asterisk}) )$ \\ \hline
\hline
$u_i u_j$ &$u_iu_l$ & $\left(\frac{m-1}{m}\right)^2$ &0\\
$u_i u_j$ &$u_ju_i$ & 0&$\frac{m-1}{m}$\\
$u_i u_j$ &$u_ju_l$ & 0& $\left(\frac{m-1}{m}\right)^2$\\
$u_i u_j$ &$u_ku_i$ & 0&$\left(\frac{m-1}{m_{\vphantom{x}}}\right)^2$\\
$u_i u_j$ &$u_ku_j$ &$\frac{m-1}{m^2}$ &$\frac{(m-1)(m-2)}{m^2}$\\
$u_i u_j$ &$u_ku_l$ & $\frac{(m-1)^2}{m^3}$ &$\frac{(m-1)^3}{m^3_{\vphantom{x}}}$\\ \hline
\end{tabular}
\end{center} 
\caption{\small Probabilities for various slot constellations.}
\end{table}

Moreover, from (\ref{eqn:X_Y}) we obtain the following decomposition:
\begin{equation}\label{eqn:decomp} \begin{aligned}\E [Y_{ij} Y_{kl}] &=  \E [X_{ij} X_{kl} \mid S_i = S_k; S_i \ne S_j; S_k \ne S_l] \\ 
&\;\,\; \cdot \Pr(S_i = S_k; S_i \ne S_j; S_k \ne S_l)\\& + \E [X_{ij} X_{kl} \mid S_i \ne S_k; S_i \ne S_j; S_k \ne S_l]\\
& \;\,\; \cdot \Pr(S_i \ne S_k; S_i \ne S_j; S_k \ne S_l)
, \end{aligned}\end{equation}
since $\Omega$ is the disjoint union of $\{\omega \in \Omega \mid S_i(\omega) = S_k(\omega)\}$ and $\{\omega \in \Omega \mid S_i(\omega) \ne S_k(\omega)\}.$

The slot probabilities can be found in Table 
II, whereas the conditioned expectations will be treated in the remaining part of this section. For instance, we have for the links $u_i u_j, u_k u_j:$ 
    \begin{equation*}\begin{aligned}\Pr&(S_j \ne S_i, S_k;\, S_i \ne S_k)\\ 
    &= \frac{1}{m^3}\sum_{b_j = 1}^m \sum_{\substack{b_i = 1\\b_i\ne b_j}}^m \sum_{\substack{b_k=1\\b_k\ne b_j,b_i}}^m 1\\
    &= \frac{(m-1)(m-2)}{m^2}.
    \end{aligned}\end{equation*}

In the following, we denote by $J$ the set $\left\{ i,j,k,l. \right\}.$

a) Let $i\ne j$ and $k \ne l.$ From Table I 
we may also assume $l\ne i$ and $k \ne j.$ Further, first let $i\ne k$ and $j\ne l.$ Then we have $\abs{J} = 4$ and 
 $$ \begin{aligned} \E&\left[ X_{ij}X_{kl} \mid S_i \ne S_j, S_k \ne S_l, S_i = S_k \right] \\	&= \Pr\left( \left.\frac{P_{ij}}{N+\sum_{\nu\ne i,j}^n{\delta_{S_i S_\nu}P_{\nu j}}}\ge \theta \mbox{ and } \right.\right.\\
	&\ds \left.\left.\frac{P_{kl}}{N+\sum_{\nu \ne k,l} \delta_{S_k S_\nu}P_{\nu l}}\ge \theta \right\rvert S_i \ne S_j, S_k \ne S_l, S_i = S_k \right) \\
	&\stackrel{\text{(a)}}{=} \frac{1}{\left( m-1 \right)^2}\frac{1}{m^{n-3}}\sum_{\substack{b_i = 1\\(b_i = b_k)}}^{m}\sum_{\substack{b_j, b_l = 1\\b_j, b_l \ne b_i}}^{m}\sum_{\substack{b_\nu = 1\\ \nu \ne i,j,k,l}}^{m} \left[\vphantom{\frac{P}{N}}\right. \\
	&\ds\ds\ds\;\;\; \cdot \Pr\left( \frac{P_{ij}}{N+\sum_{\nu\ne i,j,k,l}^{}\delta_{b_i b_\nu}P_{\nu j}+P_{kj}}\ge \theta \right)\\
	&\ds\ds\ds\;\;\; \left.\cdot \Pr\left( \frac{P_{kl}}{N+\sum_{\nu\ne k,l,i,j}^{}\delta_{b_i b_\nu}P_{\nu l}+P_{il}}\ge \theta \right)\right]\\ 
	&\stackrel{\text{(b)}}{=} \frac{1}{\left( m-1 \right)^2}\frac{1}{m^{n-3}}\sum_{\substack{b_i = 1\\(b_i = b_k)}}^{m}\sum_{\substack{b_j, b_l = 1\\b_j, b_l \ne b_i}}^{m}\sum_{\substack{c_\nu = 0\\ \nu \ne i,j,k,l}}^{1} \phi(c)\left[\vphantom{\frac{P}{N}}\right. \\
	&\ds\ds\ds\;\;\; \cdot \Pr\left( \frac{P_{ij}}{N+\sum_{\nu\ne i,j,k,l}^{}c_\nu \cdot P_{\nu j}+P_{kj}}\ge \theta \right)\\
	&\ds\ds\ds\;\;\; \left.\cdot \Pr\left( \frac{P_{kl}}{N+\sum_{\nu\ne k,l,i,j}^{}c_\nu \cdot P_{\nu l}+P_{il}}\ge \theta \right)\right]\\
	&= \frac{1}{m^{n-4}}\sum_{\substack{c_\nu = 0\\ \nu \ne i,j,k,l}}^{1} \phi(c)\left[\vphantom{\frac{P}{N}}\right.  \\
	&\ds\ds\ds\;\;\; \cdot \Pr\left( \frac{P_{ij}}{N+\sum_{\nu\ne i,j,k,l}^{}c_\nu \cdot P_{\nu j}+P_{kj}}\ge \theta \right)\\
	&\ds\ds\ds\;\;\; \left.\cdot \Pr\left( \frac{P_{kl}}{N+\sum_{\nu\ne k,l,i,j}^{}c_\nu \cdot P_{\nu l}+P_{il}}\ge \theta \right)\right],\\
\end{aligned}$$
where we have conditioned on the independent slot variables and exploited the IBF assumption in (a) and introduced macro variables in (b).
Analogous to the full-duplex variant above, the case $j=l$ leads to a contradiction and the case $i=k$ reduces the number of possible slot choices which is why we have to incorporate a prefactor $1-\delta_{jl}$ and a prefactor $m/m^{\delta_{ik}}.$ This concludes the proof for the conditioned expectation expression of part a).

The covariance part follows easily from (\ref{eqn:tot_prob}) since $S_i = S_k$ holds automatically, here.
\newpage

b) Note first that the covariance expression follows from the decomposition (\ref{eqn:decomp}) and the conditioned expectations below. Due to the $1-\delta_{lj}$ factor in the conditioned expectation of part (a), there is a contribution of the $\Phi$-expression only in case $\abs{J'} =4$ (cf. Table I).

Let $i\ne j,k$ and $k \ne l.$ 
We distinguish five cases, according to Table I. 

(i) We start with the links $u_iu_j$ and $u_k u_l$ from Table I. 
Then we find $l\ne i,j$ and $j \ne k,$ and $\abs{J}$ = 4. 
We have
	$$\begin{aligned} \E&\left( X_{ij}X_{kl} \mid S_i \ne S_j, S_k;\; S_k\ne  S_l \right) \\	&= \Pr\left( \left.\frac{P_{ij}}{N+\sum_{\nu\ne i,j}^n{\delta_{S_i S_\nu}P_{\nu j}}}\ge \theta \mbox{ and } \right.\right.\\
	&\ds \left.\left.\frac{P_{kl}}{N+\sum_{\nu \ne k,l} \delta_{S_k S_\nu}P_{\nu l}}\ge \theta \right\rvert S_i \ne S_j, S_k;\; S_k\ne  S_l\right) \\
	&\stackrel{\text{(a)}}{=} \frac{1}{ m^{n-2}}\frac{1}{m-1}\sum_{\substack{b_j,b_k = 1}}^{m}\frac{1}{m-2+\delta_{b_j b_k}}\sum_{\substack{b_i= 1\\b_i \ne b_j, b_k}}^{m}\sum_{\substack{b_l = 1\\ b_l \ne b_k}}^{m} \left[\vphantom{\frac{P}{N}}\right. \\
		&\ds\ds \sum_{\substack{b_\nu=1\\ \nu \not\in J}}^{m} \Pr\left( \left.\frac{P_{ij}}{N+\sum_{\nu\ne i,j}^n{\delta_{b_i b_\nu}P_{\nu j}}}\ge \theta \mbox{ and } \right.\right.\\ 
	&\ds\ds\ds\left.\left.\frac{P_{kl}}{N+\sum_{\nu \ne k,l} \delta_{b_k b_\nu}P_{\nu l}}\ge \theta \right)\right] \\
		\stackrel{\text{(b)}}{=} &\frac{1}{ m^{n-2}}\frac{1}{m-1}\sum_{\substack{b_j,b_k = 1}}^{m}\frac{1}{m-2+\delta_{b_j b_k}}\sum_{\substack{b_i= 1\\b_i \ne b_j, b_k}}^{m}\sum_{\substack{b_l = 1\\ b_l \ne b_k}}^{m} \left[\vphantom{\frac{P}{N}}\right. \\ 
  		& \sum_{\substack{b_\nu = 1\\ \nu \not\in J}}^{m} \Pr\left( \left.\frac{P_{ij}}{N+\sum_{\nu\ne i,j,k}^n{\delta_{b_i b_\nu}P_{\nu j}}}\ge \theta\right)\right.\\
	& \ds\ds \left.\cdot \Pr\left(\frac{P_{kl}}{N+\sum_{\nu \ne k,l,i} \delta_{b_k b_\nu}P_{\nu l}}\ge \theta \right)\right]
	\end{aligned}$$
 	$$ \begin{aligned}\stackrel{}{=} &\frac{1}{ m^{n-2}}\frac{1}{m-1}\sum_{\substack{b_j,b_k = 1}}^{m}\frac{1}{m-2+\delta_{b_j b_k}}\sum_{\substack{b_i= 1\\b_i \ne b_j, b_k}}^{m}\sum_{\substack{b_l = 1\\ b_l \ne b_k}}^{m} \left[\vphantom{\frac{P}{N}}\right. \\ 
			& \sum_{\substack{b_\nu = 1\\ \nu \not \in J}}^{m} \Pr\left( \left.\frac{P_{ij}}{N+\sum_{\nu\not\in J }^n{\delta_{b_i b_\nu}P_{\nu j}+\delta_{b_i b_l}P_{lj}}}\ge \theta\right)\right.\\
		& \ds\ds \left.\cdot \Pr\left(\frac{P_{kl}}{N+\sum_{\nu \not\in J} \delta_{b_k b_\nu}P_{\nu l}+\delta_{b_k b_j}P_{jl}}\ge \theta \right)\right]
	\end{aligned}$$
 
	\begin{equation}\label{eqn:slot_expansion} \begin{aligned} 
		\stackrel{\text{(c)}}{=} &\frac{1}{ m^{n-2}}\frac{1}{m-1}\sum_{\substack{b_j,b_k = 1}}^{m}\frac{1}{m-2+\delta_{b_j b_k}}\sum_{\substack{b_i= 1\\b_i \ne b_j, b_k}}^{m}\sum_{\substack{b_l = 1\\ b_l \ne b_k}}^{m} \left[\vphantom{\frac{P}{N}}\right. \\ 
  		& \sum_{\substack{c_\nu = 0\\ \nu \not\in J}}^{2}\psi_0(c) \cdot \Pr\left( \left.\frac{P_{ij}}{N+\sum_{\nu\not\in J}^n{\delta_{1 c_\nu}P_{\nu j}+\delta_{b_i b_l}P_{lj}}}\ge \theta\right)\right.\\
	& \ds\ds \left.\cdot \Pr\left(\frac{P_{kl}}{N+\sum_{\nu \not\in J} \delta_{2 c_\nu}P_{\nu l}+\delta_{b_k b_j}P_{jl}}\ge \theta \right)\right]
	\end{aligned}\end{equation}
	where we have conditioned on the independent slot variables in (a) and exploited the IBF assumption in (b), and introduced macro variables in (c) with state weight
		$$\psi_0(c) = \prod_{\substack{\nu = 1\\\nu \ne J}}^m (m-2)^{\delta_{c_\nu, 0}}, \ds c =  [c_\nu]_{\nu \not\in J}.$$
	
	Note that since $b_i\ne b_k,$ the active sets of interfering nodes are disjoint in (b).
Gathering the $m-2$ micro states belonging to $b_l \ne b_i$ in the lowermost expression of (\ref{eqn:slot_expansion}) yields the term
	\begin{equation}\label{eqn:isolated_term}\begin{aligned}
 		&\frac{1}{ m^{n-2}}\frac{1}{m-1}\sum_{\substack{b_j,b_k = 1}}^{m}\frac{1}{m-2+\delta_{b_j b_k}}\sum_{\substack{b_i= 1\\b_i \ne b_j, b_k}}^{m}\left[\vphantom{\frac{P}{N}}\right. \\ 
  		& \sum_{\substack{c_\nu = 0\\ \nu \not\in J}}^{2}\psi_0(c)\cdot \left[ (m-2)\cdot \Pr\left( \left.\frac{P_{ij}}{N+\sum_{\nu\not\in J }^n{\delta_{1 c_\nu}P_{\nu j}}}\ge \theta\right)\right.\right.\\
		& \ds\ds\ds\ds\ds\ds + \left.\Pr\left( \frac{P_{ij}}{N+\sum_{\nu\not\in J}^{}\delta_{1,c_\nu}P_{\nu j}+P_{lj}} \right)\right]\\
		& \ds\ds \left.\cdot \Pr\left(\frac{P_{kl}}{N+\sum_{\nu \not\in J} \delta_{2 c_\nu}P_{\nu l}+\delta_{b_k b_j}P_{jl}}\ge \theta \right)\right].
	\end{aligned}\end{equation}
The innermost summands do no longer depend on $b_i,$ thus (\ref{eqn:isolated_term}) can be written as
	$$\begin{aligned}
 		&\frac{1}{ m^{n-2}}\frac{1}{m-1}\sum_{\substack{b_j,b_k = 1}}^{m}\left[\vphantom{\frac{P}{N}}\right. \\ 
  		& \sum_{\substack{c_\nu = 0\\ \nu \not\in J}}^{2}\psi_0(c)\cdot \left[ (m-2)\cdot \Pr\left( \left.\frac{P_{ij}}{N+\sum_{\nu\not\in J}^n{\delta_{1 c_\nu}P_{\nu j}}}\ge \theta\right)\right.\right.\\
		& \ds\ds\ds\ds\ds\ds + \left.\Pr\left( \frac{P_{ij}}{N+\sum_{\nu\not\in J}^{}\delta_{1,c_\nu}P_{\nu j}+P_{lj}} \right)\right]\\
		& \ds\ds \left.\cdot \Pr\left(\frac{P_{kl}}{N+\sum_{\nu \not\in J } \delta_{2 c_\nu}P_{\nu l}+\delta_{b_k b_j}P_{jl}}\ge \theta \right)\right],
	\end{aligned}$$ 
and hence
	$$\begin{aligned} \E&\left( X_{ij}X_{kl} \mid S_i \ne S_j, S_k;\; S_k\ne  S_j \right) \\ 
		&=\frac{1}{ m^{n-2}}\frac{1}{m-1}\sum_{b_j =1}^{m}\left[\vphantom{\frac{P}{N}}\right. \\ 
  		&\ds\ds \sum_{\substack{c_\nu = 0\\ \nu \not\in J}}^{2}\psi_0(c)\cdot \left[ (m-2)\cdot \Pr\left( \left.\frac{P_{ij}}{N+\sum_{\nu\not\in J}^n{\delta_{1 c_\nu}P_{\nu j}}}\ge \theta\right)\right.\right.\\
		& \ds\ds\ds\ds\ds\ds + \left.\Pr\left( \frac{P_{ij}}{N+\sum_{\nu\not\in J}^{}\delta_{1,c_\nu}P_{\nu j}+P_{lj}} \right)\right]\\
		& \ds\ds \left.\cdot \sum_{\substack{b_k = 1}}^{m}\Pr\left(\frac{P_{kl}}{N+\sum_{\nu \not\in J } \delta_{2 c_\nu}P_{\nu l}+\delta_{b_k b_j}P_{jl}}\ge \theta \right)\right]
		\end{aligned}$$
		$$\begin{aligned}
 		&=\frac{1}{ m^{n-2}}\frac{m}{m-1}\left[\vphantom{\frac{P}{N}}\right. \\ 
  		&\ds\ds \sum_{\substack{c_\nu = 0\\ \nu \not\in J}}^{2}\psi_0(c)\cdot \left[ (m-2)\cdot \Pr\left( \left.\frac{P_{ij}}{N+\sum_{\nu\not\in J}^n{\delta_{1 c_\nu}P_{\nu j}}}\ge \theta\right)\right.\right.\\
		& \ds\ds\ds\ds\ds\ds + \left.\Pr\left( \frac{P_{ij}}{N+\sum_{\nu\not\in J}^{}\delta_{1,c_\nu}P_{\nu j}+P_{lj}} \right)\right]\\
		& \ds\ds \cdot \left[(m-1) \cdot \Pr\left(\frac{P_{kl}}{N+\sum_{\nu \not\in J } \delta_{2 c_\nu}P_{\nu l}}\ge \theta \right)\right.\\
       		& \ds\ds\ds\ds+ \left.\Pr\left(\frac{P_{kl}}{N+\sum_{\nu \not\in J } \delta_{2 c_\nu}P_{\nu l}+P_{jl}}\ge \theta \right) \right]\\
		&=\frac{1}{ m^{n-3}(m-1)}\left[\vphantom{\frac{P}{N}}\right. \\ 
  		&\ds\ds \sum_{\substack{c_\nu = 0\\ \nu \ne i,k}}^{2}\psi(c)\cdot \left[ \Pr\left( \frac{P_{ij}}{N+\sum_{\nu\ne i,j,k}^{}\delta_{1,c_\nu}P_{\nu j}} \right)\right.\\
		& \ds\ds\ds\ds\ds\ds\ds\ds\ds \left.\cdot \Pr\left(\frac{P_{kl}}{N+\sum_{\nu \ne k,l,i } \delta_{2 c_\nu}P_{\nu l}}\ge \theta \right)\right],\\
	\end{aligned}$$                                                                                                                                   
where, for $c =  [c_\nu]_{\nu \ne i,k},$ 
$$\begin{aligned}\psi(c) &= \delta_{c_l\ne 2}\cdot \delta_{c_j\ne 1}\cdot ( m-1)^{\delta_{0,c_j}}(m-2)^{\delta_{c_l,0}}\cdot\psi_0([c_\nu]_{\nu \not\in J})\\
	&= \delta_{c_l\ne 2}\cdot \delta_{c_j\ne 1}\cdot ( m-1)^{\delta_{0,c_j}}\cdot\prod_{\substack{\nu = 1\\ \nu \ne i,j,k}}^n (m-2)^{\delta_{c_\nu, 0}}\end{aligned}.$$ 

(ii) The links $u_i u_j$ and $u_k u_j$ represent the case $j\ne k, j = l.$ Similarly, we get
 	$$\begin{aligned} \E&\left( X_{ij}X_{kj} \mid S_i \ne S_j, S_k;\; S_k\ne  S_l \right) \\	=&\, \Pr\left( \left.\frac{P_{ij}}{N+\sum_{\nu\ne i,j}^n{\delta_{S_i S_\nu}P_{\nu j}}}\ge \theta \mbox{ and } \right.\right.\\
	&\ds \left.\left.\frac{P_{kj}}{N+\sum_{\nu \ne k,j} \delta_{S_k S_\nu}P_{\nu j}}\ge \theta \right\rvert S_i \ne S_j, S_k;\; S_k\ne  S_j\right) \\ 
 	\stackrel{}{=} &\frac{1}{ m^{n-2}}\frac{1}{m-1}\sum_{\substack{b_j,b_k = 1\\b_j\ne b_k}}^{m}\frac{1}{m-2}\sum_{\substack{b_i= 1\\b_i \ne b_j, b_k}}^{m}\left[\vphantom{\frac{P}{N}}\right. \\ 
  		& \sum_{\substack{b_\nu = 0\\ \nu \not\in J}}^{2} \Pr\left( \left.\frac{P_{ij}}{N+\sum_{\nu\ne i,j,k}^n{\delta_{b_i b_\nu}P_{\nu j}+0}}\ge \theta\right)\right.\\
	& \ds\ds \left.\cdot \Pr\left(\frac{P_{kj}}{N+\sum_{\nu \ne k,j,i} \delta_{b_k b_\nu}P_{\nu j}+0}\ge \theta \right)\right]
	\end{aligned}$$
	$$\begin{aligned}
 	\stackrel{}{=} &\frac{1}{ m^{n-2}}\frac{1}{m-1}\sum_{\substack{b_j,b_k = 1\\b_j\ne b_k}}^{m}\frac{1}{m-2}\sum_{\substack{b_i= 1\\b_i \ne b_j, b_k}}^{m}\left[\vphantom{\frac{P}{N}}\right. \\ 
  		& \sum_{\substack{c_\nu = 0\\ \nu \not\in J}}^{2}\psi(c) \cdot \Pr\left( \left.\frac{P_{ij}}{N+\sum_{\nu\not\in J }^n{\delta_{1 c_\nu}P_{\nu j}}}\ge \theta\right)\right.\\
	& \ds\ds \left.\cdot \Pr\left(\frac{P_{kj}}{N+\sum_{\nu \not\in J} \delta_{2 c_\nu}P_{\nu j}}\ge \theta \right)\right]\\
 	\stackrel{}{=} &\frac{1}{ m^{n-2}}\frac{1}{m-1}\sum_{\substack{b_j,b_k = 1\\b_j\ne b_k}}^{m}\left[\vphantom{\frac{P}{N}}\right. \\ 
  		& \sum_{\substack{c_\nu = 0\\ \nu \not\in J}}^{2}\psi(c) \cdot \Pr\left( \left.\frac{P_{ij}}{N+\sum_{\nu\not\in J }^n{\delta_{1 c_\nu}P_{\nu j}}}\ge \theta\right)\right.\\
	& \ds\ds \left.\cdot \Pr\left(\frac{P_{kj}}{N+\sum_{\nu \not\in J } \delta_{2 c_\nu}P_{\nu j}}\ge \theta \right)\right]  \\
 	\stackrel{}{=} &\frac{1}{ m^{n-3}}\left[\vphantom{\frac{P}{N}}\right. \\ 
  		& \sum_{\substack{c_\nu = 0\\ \nu \not\in J}}^{2}\psi(c) \cdot \Pr\left( \left.\frac{P_{ij}}{N+\sum_{\nu\not\in J }^n{\delta_{1 c_\nu}P_{\nu j}}}\ge \theta\right)\right.\\
	& \ds\ds \left.\cdot \Pr\left(\frac{P_{kj}}{N+\sum_{\nu \not\in J } \delta_{2 c_\nu}P_{\nu j}}\ge \theta \right)\right]  \\ 
\end{aligned}$$ 
where, for $c =  [c_\nu]_{\nu \not\in J},$
	$$\psi(c) = \prod_{\substack{\nu = 1\\ \nu \not\in J}}^n (m-2)^{\delta_{c_\nu, 0}}.$$
(iii) We come to the links $u_i u_j$ and $u_k u_i$ representing the case $l=i,\; j \ne k.$ This case is shown analogous to the previous case (ii) and is thus is omitted. 

(iv) The links $u_i u_j$ and $u_j u_l$ represent the case $k=j,\; l \ne i.$ We have
 	$$\begin{aligned} \E&\left( X_{ij}X_{jl} \mid S_j \ne S_i, S_l \right) \\	
 	\stackrel{}{=} &\frac{1}{ m}\frac{1}{(m-1)^2}\sum_{\substack{b_j= 1}}^{m}\sum_{\substack{b_i= 1\\b_i \ne b_j}}^{m}\sum_{\substack{b_l= 1\\b_l \ne b_j}}^{m}\left[\vphantom{\frac{P}{N}}\right. \\ 
		& \frac{1}{m^{n-3}}\sum_{\substack{b_\nu = 0\\ \nu \not\in J}}^{2} \Pr\left( \left.\frac{P_{ij}}{N+\sum_{\nu\ne i,j}^n{\delta_{b_i b_\nu}P_{\nu j}}}\ge \theta\right)\right.\\
	& \ds\ds \left.\cdot \Pr\left(\frac{P_{jl}}{N+\sum_{\nu \ne j,l,i} \delta_{b_k b_\nu}P_{\nu l}}\ge \theta \right)\right]\\ 
	\stackrel{}{=} &\frac{1}{m^{n-2}(m-1)^2}\sum_{\substack{b_j= 1}}^{m}\sum_{\substack{b_i= 1\\b_i \ne b_j}}^{m}\sum_{\substack{b_l= 1\\b_l \ne b_j}}^{m}\left[\vphantom{\frac{P}{N}}\right. \\ 
  		& \sum_{\substack{b_\nu = 0\\ \nu \not\in J}}^{2} \Pr\left( \left.\frac{P_{ij}}{N+\sum_{\nu\ne i,j}^n{\delta_{b_i b_\nu}P_{\nu j}}}\ge \theta\right)\right.\\
  		\end{aligned}$$
 	$$\begin{aligned}
	& \ds\ds \left.\cdot \Pr\left(\frac{P_{jl}}{N+\sum_{\nu \ne j,l,i} \delta_{b_k b_\nu}P_{\nu l}}\ge \theta \right)\right]\\
	\stackrel{}{=} &\frac{1}{m^{n-2}(m-1)^2}\sum_{\substack{b_j= 1}}^{m}\sum_{\substack{b_i= 1\\b_i \ne b_j}}^{m}\sum_{\substack{b_l= 1\\b_l \ne b_j}}^{m}\left[\vphantom{\frac{P}{N}}\right. 
 	\\
 	& \sum_{\substack{c_\nu = 0\\ \nu \not\in J}}^{2} \psi(c) \cdot \Pr\left( \left.\frac{P_{ij}}{N+\sum_{\nu\ne i,j}^n{\delta_{c_\nu, 1}P_{\nu j}}}\ge \theta\right)\right.\\
	& \ds\ds \left.\cdot \Pr\left(\frac{P_{jl}}{N+\sum_{\nu \ne j,l,i} \delta_{c_\nu,2}P_{\nu l}}\ge \theta \right)\right]\\   
  	\stackrel{}{=} &\frac{1}{m^{n-3}}\left[\sum_{\substack{c_\nu = 0\\ \nu \not\in J}}^{2} \psi(c) \cdot \Pr\left( \left.\frac{P_{ij}}{N+\sum_{\nu\ne i,j}^n{\delta_{c_\nu, 1}P_{\nu j}}}\ge \theta\right)\right.\right.\\
	& \ds\ds \left.\cdot \Pr\left(\frac{P_{jl}}{N+\sum_{\nu \ne j,l,i} \delta_{c_\nu,2}P_{\nu l}}\ge \theta \right)\right],\\    
	\end{aligned}$$
where 
 here, for $c =  [c_\nu]_{\nu \ne J},$
	$$\psi(c) = \prod_{\substack{\nu = 1\\ \nu \ne i,j,l}}^n (m-2)^{\delta_{c_\nu, 0}}.$$ 
(v) Finally, the links $u_i u_j$ and $u_j u_i$ represent the case $k = j$ and $l = i.$ 
 	$$\begin{aligned} \E&\left( X_{ij}X_{ji} \mid S_i \ne S_j, S_k \right) \\
 		\stackrel{}{=} &\frac{1}{ m}\frac{1}{m-1}\frac{1}{m^{n-2}}\sum_{\substack{b_j=1}}^{m}\sum_{\substack{b_i= 1\\b_i \ne b_j}}^{m}\left[\vphantom{\frac{P}{N}}\right. \\ 
  		& \sum_{\substack{b_\nu = 0\\ \nu \ne i,j}}^{2} \Pr\left( \left.\frac{P_{ij}}{N+\sum_{\nu\ne i,j}^n{\delta_{b_i b_\nu}P_{\nu j}}}\ge \theta\right)\right.\\
	& \ds\ds \left.\cdot \Pr\left(\frac{P_{kj}}{N+\sum_{\nu \ne j,i} \delta_{b_k b_\nu}P_{\nu j}}\ge \theta \right)\right]\\  
 		\stackrel{}{=} &\frac{1}{ m}\frac{1}{m-1}\frac{1}{m^{n-2}}\sum_{\substack{b_j=1}}^{m}\sum_{\substack{b_i= 1\\b_i \ne b_j}}^{m}\left[\vphantom{\frac{P}{N}}\right. \\ 
  		& \sum_{\substack{c_\nu = 0\\ \nu \ne i,j}}^{2} \psi(c)\cdot\Pr\left( \left.\frac{P_{ij}}{N+\sum_{\nu\ne i,j}^n{\delta_{c_\nu,1}P_{\nu j}}}\ge \theta\right)\right.\\
	& \ds\ds \left.\cdot \Pr\left(\frac{P_{kj}}{N+\sum_{\nu \ne j,i} \delta_{c_\nu,2}P_{\nu j}}\ge \theta \right)\right]\\   
 		\stackrel{}{=} &\frac{1}{m^{n-2}}\left[\sum_{\substack{c_\nu = 0\\ \nu \ne i,j}}^{2} \psi(c)\cdot\Pr\left( \frac{P_{ij}}{N+\sum_{\nu\ne i,j}^n{\delta_{c_\nu,1}P_{\nu j}}}\ge \theta\right)\right.\\
	& \ds\ds\ds\ds\ds\ds \left.\cdot \Pr\left(\frac{P_{kj}}{N+\sum_{\nu \ne j,i} \delta_{c_\nu,2}P_{\nu j}}\ge \theta \right)\right],\\     	
\end{aligned}$$  
where, for $c =  [c_\nu]_{\nu \ne i,j},$
	$$\psi(c) = \prod_{\substack{\nu = 1\\ \nu \ne i,j}}^n (m-2)^{\delta_{c_\nu, 0}}.$$
This concludes the proof of part b) and thus of Theorem \ref{prp:slotCorrelationsHD}.
\qed 

\subsection{Proof of Corollary \ref{cor:nakagamiHD}}
\label{proof:cor_HD}
We show that by \cite{Nakagami12} the identity
\begin{equation}\label{eqn:gamma_ij}\gamma_{ij}^{\mathcal N\setminus \left\{ i,j \right\}}([\xi_\nu]_{\nu\in \mathcal N\setminus\left\{ i,j \right\}}) = \Pr\left(\frac{P_{ij}}{N+\sum\limits_{\nu\ne i,j}^{}\Xi_\nu \cdot P_{\nu j}} \ge \theta\right)\end{equation}
holds for Bernoulli variables $\Xi_\nu$ with parameters $\xi_\nu$ such that $\Xi_\nu, P_{ij},$ and $P_{\nu, j},$ for $\nu \ne i,j,$ are mutually independent. 
We verify that $\Xi_{\nu} := \delta_{S_i S_{\nu}}$ satisfies this property such that then $\E X_{ij} = \gamma_{ij}(\one/m).$ An application of (\ref{eqn:expectation_decomp}) then establishes the first part of the corollary.

The second part of the corollary then follows analogously by using only probabilities $\xi_\nu = \xi_\nu \in \{0,1\}$ and from Theorem \ref{prp:slotCorrelationsHD}. 
We divide the proof into three sections. 

(i) We start with the verification of the independence property. For $\Xi_\nu = \delta_{S_i S_{\nu}}$ and $x,y \in \left\{ 0,1 \right\},\; \nu \ne i,j$ we have 

\begin{equation*}
	\begin{aligned}
		\Pr(\Xi_\nu &= x_\nu, \; \nu \ne i,j)\\
		&= \frac{1}{m}\sum_{b_i = 1}^{m}\Pr(\delta_{b_i S_\nu} = x_\nu , \; \nu \ne i,j \mid S_i = b_i)\\
		&= \frac{1}{m}\sum_{b_i = 1}^{m}\Pr(\delta_{b_i S_\nu} = x_\nu , \; \nu \ne i,j )\\
		&= \Pr(\delta_{1 S_\nu} = x_\nu , \; \nu \ne i,j )\\
		&= \prod_{\substack{\nu =1\\ \nu \ne i,j}}^n \Pr\left( \delta_{1 S_\nu} = x_\nu \right)\\
		&= \prod_{\nu \ne i,j} \left( \frac{1}{m} \sum_{b_i = 1}^{m} \Pr( \delta_{b_i S_\nu} = x_\nu \mid S_i = b_i   \right)\\
		&= \prod_{\nu \ne i,j} \Pr(\Xi_\nu = x_\nu),\\
	\end{aligned}
\end{equation*} 
where we several times made use of the independence and the uniformness of the slot choices. This shows the mutual independence of $\Xi_\nu,\, \nu \ne i,j.$  The remaining cases are immediately clear from the IBF assumption and the independent choice of slots.

(ii) We now establish the connection (\ref{eqn:gamma_ij}). First note that the transmitter's index 0 of \cite{Nakagami12} corresponds to the index $i$ here. Since the transmitter does not represent a designated node in our case, we rearrange the indexing in \cite{Nakagami12} to fit our notation: $\Omega_i$ corresponds to the transmitter, $\Omega_j$ to the receiver and $\Omega_\nu, \nu = 1, \dots, n, \; \nu \ne i,j$ to the interferers.
We set
	$$\Omega_\nu = \frac{1}{q_{i}\cdot \mu_{\nu j}}, \; \nu \ne j, $$
where we used unity reference distance ($d_0 = 1$) since our reference distance $r_0$ is already incorporated in the $\mu_{\nu j}$'s as well as the $q_\nu$'s, the transmit powers. Further, we set
	$$\begin{aligned}
		\Gamma &= \frac{q_{i}}{N}, \ds\ds\ds\ds
	&
		&g_{\nu}= \mu_{\nu j} \cdot P_{\nu j},\ds \nu \ne j,
	\\
		 \beta &= \theta,  \ds\ds\ds\ds
	&
		 &S = \frac{\Omega_i g_i}{\beta},
	\\
		 \beta_i &= \beta \frac{m_i}{\Omega_i}, \ds\ds\ds\ds
	&
		&Y_\nu = \Xi_\nu \cdot \Omega_\nu \cdot g_\nu, \ds \nu \ne j,
	\\
 	\end{aligned}
	$$ 
Again the index $i$ corresponds to the transmitter, $j$ to the receiver while all other indices represent interferers. Then $g_\nu,\, \nu \ne j,$ is a Nakagami fading variable with shape $m_\nu$ and unit average power at the receiver $j.$ Therefore, we have

$$\begin{aligned}
 	&\Pr\left(\frac{P_{ij}}{N+\sum\limits_{\nu\ne i,j}^{}\Xi_\nu \cdot P_{\nu j}} \ge \theta\right)\\
	&=\Pr\left(\frac{\mu_{ij}^{-1} g_i}{N+\sum\limits_{\nu\ne i,j}^{}\Xi_\nu \cdot \mu_{\nu j}^{-1} g_\nu} \ge \theta\right)\\
	&=\Pr\left(\frac{\mu_{ij}^{-1} g_i/q_i}{N/q_i+\sum\limits_{\nu\ne i,j}^{}\Xi_\nu \cdot \mu_{\nu j}^{-1} g_\nu/q_i} \ge \beta\right)\\
	&=\Pr\left( \frac{\Omega_i g_i}{\Gamma^{-1} + \sum_{\nu \ne i,j}^{}Y_\nu} \ge \beta\right)\\
	&=\Pr\left( \Gamma^{-1} \le S -  \sum_{\nu \ne i,j}^{}Y_\nu \right)\\
	&\stackrel{\text{(a)}}{=} e^{-\beta_i\cdot \Gamma^{-1}} \sum_{s=0}^{m_{i}-1}\left(\beta_i \Gamma^{-1} \right)^s \\& \cdot \sum_{t=0}^{s}\frac{\Gamma^t}{\left( s-t \right)!}   \cdot \sum_{\substack{\ell_\nu \ge 0\\ \sum\limits_{\nu \ne i,j}\ell_\nu = t}}\prod_{\nu \ne i,j}\left[ \vphantom{\frac{a^2}{b^2}}\left( 1-\xi_{\nu} \right)\delta_{0\ell_\nu} \right. \\ &\ds\ds +\left. \frac{\xi_{\nu}\cdot \left( \ell_\nu + m_\nu -1\right)!\left( \frac{\Omega_\nu}{m_\nu } \right)^{\ell_\nu}}{\ell_{\nu}! \cdot (m_\nu-1)!\left( \beta_i \frac{\Omega_i}{m_i}+1 \right)^{m_\nu+\ell_\nu} }  \right]. \\
	 &= e^{-\theta\cdot m_{i} \cdot \mu_{ij} N} \sum_{s=0}^{m_{i}-1}\left( \theta \cdot m_{i} \cdot \mu_{ij} N \right)^s \\& \cdot \sum_{t=0}^{s}\frac{(\frac{q_{i}}{N})^t}{\left( s-t \right)!}   \cdot \sum_{\substack{\ell_\nu \ge 0\\ \sum\limits_{\nu \ne i,j}\ell_\nu = t}}\prod_{\nu \ne i,j}\left[ \vphantom{\frac{a^2}{b^2}}\left( 1-\xi_{\nu} \right)\delta_{0\ell_\nu} \right. 
	\\
 &\ds\ds +\left. \left(\begin{aligned} \ell_\nu + &m_\nu -1\\ &\ell_{\nu}\end{aligned} \right)\cdot\frac{\xi_{\nu}\cdot \left( \frac{1}{m_\nu q_{i} \mu_{\nu j}} \right)^{\ell_\nu}}{\left( \theta\frac{m_i}{m_\nu}\frac{\mu_{ij}}{\mu_{\nu j}}+1 \right)^{m_\nu+\ell_\nu} }  \right]\\
&= \gamma_{ij}^{\mathcal N\setminus \left\{ i,j \right\}}([\xi_\nu]_{\nu\in \mathcal N\setminus\left\{ i,j \right\}}), 
\end{aligned}
$$
where (a) follows from \cite[eqn. (11), (12) and (24)]{Nakagami12}. Together with (\ref{eqn:expectation_decomp}) 
 this yields
	$$\E Y_{ij} = \left(1-\frac{1}{m}\right) \cdot \gamma_{ij}^{\mathcal N\setminus \{i,j\}}(\one/m)$$
since 
$$\E \delta_{S_i S_\nu} = \frac{1}{m^2}\cdot \sum_{b_i,b_\nu = 1}^{m} \delta_{b_i b_\nu} = \frac{1}{m}.$$

(iii) The state coefficients of Theorem \ref{prp:slotCorrelationsHD} can now be expressed in virtue of (\ref{eqn:gamma_ij}) which concludes the proof.
\qed
 
\subsection{Proof of Theorem 
\ref{thm:bounds}}
\label{prf:thm:bounds}

First let $W = Z.$ Let us check that $R^Z(\varepsilon)$ equals the matrix $\tilde R$ of \cite{ICCCN19} (hence $w_2(Z, \varepsilon)$ corresponds  to $\tilde w_2(L,\varepsilon)$) of \cite{ICCCN19}. Indeed, using the notation $P^Z = (I_n- \varepsilon L^Z),$ the calculus for the Kronecker product ($(AB)\otimes (AB) = (A \otimes A)\cdot(B\otimes B)$) and linearity of the expectation, we have 
	$$\tilde R := \E [(\Pi P^Z)\otimes(\Pi P^Z)] = \left( \Pi \otimes \Pi \right) \cdot \E [(P^Z\Pi)\otimes(P^Z\Pi)],$$ 
since 
	$$\Pi P^Z = \Pi P^Z \Pi,$$
which is due to $L^Z\ones = 0.$
Further, 
	$$
\begin{aligned}
	\E [(P^Z\Pi)\otimes(P^Z\Pi)] &= \E [ \Pi \otimes \Pi - \varepsilon \Pi \otimes L^Z \\
		\ds \ds &- \varepsilon L^Z \otimes \Pi + \varepsilon^2 L^Z \otimes L^Z ] \\
		&= \left( \E P^Z \Pi \right) \otimes \left( \E P^Z \Pi \right) + \varepsilon^{2} C,
\end{aligned}
$$
where $C := \E [L^Z \otimes L^Z] - \E L^Z \otimes \E L^Z.$ Clearly, the components of $C$ are given by 
$C_{IJ} = \cov(\ell_{ij}^Z, \ell_{kl}^Z),$ for $I = (n-1)\cdot i +k , J = (n-1)\cdot j + l, \ds i,j,k,l = 1 ,\dots, n.$ 
Hence $\tilde R = R^Z(\varepsilon).$

Second, we will need the following matrix norms: For any matrix $A\in \R^{n\times n}$ define the spectral norm of $A$ by $\n{A}_2 := \sqrt{\varrho(A^T A)}$ and the Frobenius norm of $A$ by 
$\n{A}_F := \sqrt{\sum_{i=1}^{n}\n{Ae_i}^2},$ 
where $e_i\in \R^n, i = 1, \dots, n$ denote the Cartesian basis vectors, i.e. $e_i = (0, \dots, 0, 1, 0, \dots 0),$ where the one is located at the $i$-th position.
Note that the Frobenius norm upper bounds the spectral norm. Analogous to the proof of \cite[Prop. 4]{ICCCN19} we define 
	$$S(X) = \E (\Pi P^Z)^T X (\Pi P^Z), \ds X \in \R^{n\times n},$$
	and utilize the inequality (cf. \cite{ICCCN19})
\begin{equation}\label{eqn:Frobenius_inequaltiy} \n{S^k(X)}_F \le \n{\tilde{R}^k}_2 \cdot \n{X}_F, \ds k\in \N. \end{equation}
Note that 
\begin{equation}\label{eqn:delta_prop}\delta_k(x) = \Pi (P_k \dots P_1)x = (\Pi P_k) \dots (\Pi P_1)x\end{equation}
for a sequence of independent variables $P_i, i= 1,2, \dots$ with $P_i \stackrel{\text{d}}{=} P^Z$ and for $x\in \R^n.$
Then we have
\begin{equation}
\label{eqn:appendix_upper_bound}
\begin{aligned}
	\max_{\n{x}\le 1} \E \delta_k^2(x) &= \max_{\n{x}\le 1}\E \n{\Pi (P_k \dots P_1)x}^2 \\
         &= \max_{\n{x}\le 1}x^T \E(P_1^T \dots P_k^T) \Pi (P_k \dots P_1)x \\ 
	 &\stackrel{\text{(a)}}{=} \n{\E(P_1^T \dots P_k^T) \Pi (P_k \dots P_1)}_2\\
	 &\stackrel{\text{}}{\le} \n{\E(P_1^T \dots P_k^T) \Pi (P_k \dots P_1)}_F\\
	 &= \n{S^k(\Pi)}_F \\
	 &\stackrel{\text{(b)}}{\le} \sqrt{n-1}\cdot \n{\tilde R^k}_2 \\
	 &\stackrel{\text{(c)}}{\le} 2 \sqrt{n-1}\cdot w_2^{2k}(Z,\varepsilon). \\
\end{aligned}
\end{equation}
where (a) is due to a property of the spectral norm and in (b) we have used (\ref{eqn:Frobenius_inequaltiy}) and the fact that $\n{\Pi}_F = \sqrt{n-1}.$  For (c) cf. the properties of the numerical radius, e.g. \cite{ICCCN19}, \cite{He1997a}. This concludes the proof for the right-hand inequality, the upper bound.

The left-hand inequality follows from the infimum property of $r_2$ 
(i.e. $r_2^{ 2k } = \inf_{k\in \N} \E \n{\Pi P_k \dots P_1}_F^2$, e.g \cite[Lem. 2.7]{Ogura2013}):
\begin{equation}
\begin{aligned}
	\label{eqn:appendix_lower_bound}
	\frac{1}{n}r^{2k}(Z,\varepsilon) \le \frac{1}{n} \E \n{\Pi P_k \dots P_1}_F^2 &\le \frac{1}{n} \E \sum_{i=1}^{n}\delta_k^2(e_i)\\
	&\le \max_{\n{x}\le 1} \E \delta_k^2(x),
\end{aligned}
\end{equation}
where we have used (\ref{eqn:delta_prop}) and the definition of the Frobenius norm which concludes the proof of the case $W= Z.$


The proof directly carries over to the case $W = Y,$ (and the case $W = X$ as well), the only difference being the structure of the matrix $C$ which is not relevant in proofing the estimates.

\subsection{Details on Remark \ref{rem:thm_bounds}}
\label{sec:appendix_additional}
Here we we provide additional information for Theorem \ref{thm:bounds} by supplementing Remark \ref{rem:thm_bounds}:
\begin{abc}
\item Let $W = Y$ or $W = Z.$ Then we have from (\ref{eqn:appendix_lower_bound}) and (\ref{eqn:appendix_upper_bound})
$$\sqrt[2k]{\frac{1}{n}r_2^{2k}(W)} \le \sqrt[2k]{\max_{\n{x}\le 1} \E \delta_k^2(x)} \le \sqrt[2k]{\sqrt{n-1}\n{\left(R^W\right)^k}_2}$$
which gives in the limit, using the spectral radius formula,
	$$r_2(W, \varepsilon) \le \lim_{k\sra \infty}\sqrt[2k]{\max_{\n{x}\le 1} \E \delta_k^2(x)} \le \sqrt{\varrho\left(R^W\right)} = r_2(W, \varepsilon).$$

\item The required steps increase logarithmically as $n$ grows larger: Let $\delta > 0.$ Then
		$$\sqrt[2k]{2 \sqrt{n-1}} < 1+\delta$$
if and only if
		$$k > \frac{1}{2}\log_{(1+\delta)}(2)+\frac{1}{4}\log_{(1+\delta)}(n-1).$$

\item From (\ref{eqn:appendix_upper_bound}) we have using $k=1$

	$$\max_{\n{x}\le 1} \E \delta_1^2(x) = \n{\E P_1^T \Pi P_1}_2$$
which is the bound used in \cite{Silva11}.
\end{abc}

\fi

\end{document}